\documentclass[preprint,showpacs,floats,floatfix,nofootinbib]{revtex4-1}

\usepackage[tbtags]{amsmath}
\usepackage{amsthm,amssymb,graphicx,mathtools,tikz,hyperref,bbm}
\usepackage{bm}
\usepackage{cleveref}

\input{epsf}
\hypersetup{
 colorlinks,
 linkcolor=blue,
 citecolor=red,
 urlcolor=blue
 }

\usepackage{xcolor}
\usepackage{textcomp}

\def\veps{{\varepsilon}}

\newcommand{\A}{\ensuremath{\mathcal{A}}}
\newcommand{\B}{\ensuremath{\mathcal{B}}}
\newcommand{\M}{\ensuremath{\mathcal{M}}}
\newcommand{\N}{\ensuremath{\mathcal{N}}}

\def\x{{\rm x}}
\def\y{{\rm y}}
\def\s{{\rm s}}
\def\n{{\rm n}}
\def\e{{\rm e}}
\def\o{{\rm o}}
\def\f{{\rm f}}

\def\d{{\rm d}}
\def\b{{\rm b}}

\def\ha{{\hat a}}
\def\hb{{\hat b}}
\def\hc{{\hat c}}
\def\hd{{\hat d}}
\def\hi{{\hat i}}

\def\fint{\int_{\cal M}}
\def\tintp{\int_{\partial {\cal M}_+}}
\def\tintm{\int_{\partial {\cal M}_-} }
\def\tintL{\int_{\partial {\cal M}_L} }

\begin{document}\count\footins = 1000
\title{Linearizing a Non-linear Formulation for General Relativistic Dissipative Fluids}

\author{T. Celora}
\thanks{T.Celora@soton.ac.uk}
\affiliation{Mathematical Sciences and STAG Research Centre, University of Southampton, Southampton SO17 1BJ, UK}
\author{N. Andersson}
\affiliation{Mathematical Sciences and STAG Research Centre, University of Southampton, Southampton SO17 1BJ, UK}
\author{G.L. Comer}
\affiliation{Department of Physics, Saint Louis University, St. Louis, MO 63156-0907, USA}

\date{\today}

\begin{abstract}
Fully non-linear equations of motion for dissipative general relativistic multi-fluids can be obtained from an action principle involving the explicit use of lower dimensional matter spaces. More traditional strategies for incorporating dissipation---like the famous M\"uller-Israel-Stewart model---are based on expansions away from equilibrium defined, in part, by the laws of thermodynamics. The goal here is to build a formalism to facilitate comparison of the action-based results with those based on the traditional approach. The first step of the process is to use the action-based approach itself to construct self-consistent notions of equilibrium. Next, first-order deviations are developed directly on the matter spaces, which motivates the latter as the natural arena for the underlying thermodynamics. Finally, we identify the dissipation terms of the action-based model with first-order ``thermodynamical'' fluxes, on which the traditional models are built. The description is developed in a general setting so that the formalism can be used to describe multi-fluid systems, for which causal and stable models are not yet available. As an illustration of the approach, a simple application of a single viscous fluid is considered and, even though the expansion is halted at first order, we sketch how a causal response can be implemented through Cattaneo-type equations.

\end{abstract}

\maketitle

\section{Introduction}
The covariant nature of general relativity highlights the central role played by the reference frame used to describe a physical system. At the same time, the evolution of dissipative fluids must be consistent with  thermodynamical principles and the arrow of time associated with the second law. Matching these two pictures---general relativity and thermodynamics---poses interesting foundational questions, and therefore, it is not surprising that the construction of general relativistic models for dissipative fluids constitutes a problem that has kept physicists busy for a long time. 

Early attempts, like the seminal work of \citet{Eckart} and \citet{LandauFLuidMechanics}, date back to the first half of the last century. These models are essentially the same and---even though the Landau-Lifschitz version is a little less pathological---have been proven to suffer from causality and stability shortcomings, neither of which can be ignored in a relativistic approach. 

Important steps forward were taken by M\"ueller \cite{Muller:1967zza} in the 1960s and Israel and Stewart in the 1970s \cite{ISRAEL1976,transientStewart77,IsraelStewart79,IsraelStewart79bis}. Their results have been shown (\citet{HiscockLindblom1983}) to resolve the stability and causality issues of the earlier attempts. However, a number of other issues remain to be addressed. 
First, the M\"uller-Israel-Stewart (MIS) model is based on an implicit expansion in deviations away from thermal equilibrium, and has been demonstrated to become unstable when large deviations are considered (see \citet{Hiscock88NonlinearPathologies}). More importantly, the equations of motion (EoM) are obtained from the conservation of the total stress-energy-momentum tensor of the system, and it is not clear how to extend the model to multi-fluid systems relevant in, say, astrophysics and cosmology.

Another important step, at least from the formal point of view, was taken by Carter \cite{CarterDissipativeModel}. His model is based on a variational principle in which thermodynamic fluxes are upgraded to dynamical variables. However, in this approach the action is only used to obtain the structure of the force terms, and the stress-energy-momentum conservation as a Noether identity. The final equations of motion are not obtained by setting to zero the variation of the action. Moreover, in order to complete the identification of the new dynamical fields with the usual thermodynamical fluxes, a specific expansion in deviations from thermal equilibrium had to be introduced, the resulting model was shown to belong to the same family as those of the MIS variety (see \cite{Priou1991} for a detailed comparison).

More recently, a fairly general procedure for deriving the field equations for dissipative fluids from an action principle has been put forward by \citet{2015CQAnderssonComer}. It extends the convective variational principle for perfect multi-fluids introduced by \citet{CarterNoto} to include dissipation---notably, maintaining the particle fluxes as the only dynamical fields. The approach is ``fully variational'' in the sense that the final equations of motion are obtained as Euler-Lagrange equations starting from an action. This feature makes the model well suited for describing dissipative multi-fluids\footnote{As opposed to simple fluids with, possibly, heat flow, which the standard approaches are designed for.}.
Moreover, the action and the field equations it produces are fully non-linear.  A familiar example illustrating this same feature is the variational principle for the Einstein-Hilbert action for General Relativity, which yields the Einstein Equations, that are notoriously non-linear in the metric. 
The key point is that, unlike the MIS approach, the \citet{2015CQAnderssonComer} action principle does not reference any sort of chemical, dynamical, or thermal equilibrium, other than to start from the assumption that the physics can be modelled as fluid phenomena. 

The main goal of the present work is to work out the linear limit of the action based model and facilitate a comparison with previous approaches (such as MIS). MIS---and more recent works \cite{Kovtun19,Bemfica19}---explicitly use an expansion to create an approximate set of field equations to describe dissipative phenomena. 
Because the action-based model already provides a set of equations (at least in principle) valid in every regime, we can carry out the analysis using standard perturbation techniques. The dissipation terms are assumed to generate first-order deviations away from equilibria obtained using the non-dissipative limit of the field equations. Working this way we hope to also understand better the impact of length- and time-scales of fluid elements on the large scale behavior of the system; in particular, how to link the micro-scale dynamics of the many particles in a fluid element with the macro-scale dynamics between the fluid elements themselves, and the role of the Equivalence Principle in setting these scales.

The paper is laid out as follows: In \cref{sec:actionbased} we briefly summarize the action-based model. In \cref{sec:thermoEq} we discuss the role of equilibrium from the fluid perspective and study the dissipative action-based equations in this limit. In \cref{sec:PerturbativeExpansion} we set the stage for the perturbative expansion around equilibrium, building it directly on the matter space. We also introduce a conceptual novelty, the ``equilibrium observer'' frame of reference, to be distinguished from the Landau or Eckart frames, and show how it arises naturally (as a diffeomorphism) in the formalism. In \cref{sec:EnergyMinimized} we impose the (thermodynamically motivated) condition that energy (density) is minimized at equilibrium and show that the dissipative contributions to the field equations vanish in this limit. This is a novel result because this behaviour is normally assumed. In \cref{sec:LastPiece,sec:StdQuantities} we complete the expansion and identify the thermodynamical fluxes in the variational context. In \cref{sec:ViscousFluid} we provide an explicit example by applying the formalism to a single viscous fluid; this also provides context for discussing the key issue of causality. Finally, in \cref{sec:Cattaneo}, we sketch how one may implement a causal response by using equations of the Cattaneo-type for the fluxes. We show that this can be done at first order in the action-based model, retaining, in realistic physical applications, the compatibility with the second law of thermodynamics.

Before moving on, it is useful to clarify the notation used throughout the discussion. We use latin letters $a,b,c,\dots$ for spacetime indices, and roman letters---such as $\x,\y$ (or $\n,\s$ in specific models)---to label different chemical species and to distinguish dissipative ($\d$) from equilibrium ($\e$) processes. Capital letters $A,B,C,\dots$ are used for matter-space indices (see below). Note that the Einstein summation convention does not apply to the chemical indices.

\section{Action-based approach to dissipative fluids: a brief recap}\label{sec:actionbased}
In order to establish the general framework, we begin by outlining the action-based approach to modelling dissipative fluids of \cite{2015CQAnderssonComer}, while referring to \cite{NilsGregReview} for an extensive review of the general variational strategy for modelling (multi-)fluids in general relativity. In the case of a (multi-)fluid system the fundamental (matter) fields can be taken as the particle four-currents. Therefore, denoting the particle fluxes of the various species\footnote{Hereafter we will not make a distinction between the words species and constituents. For instance, in a neutron star context,  the protons and the electrons constitute  examples of chemical species/constituents. When considered individually, it would be natural to think of the particle species, but when the two are locked together (leading to a charge-neutral conglomerate) the word constituent may be more appropriate. However, we do not need to make this distinction for the present discussion.} as  $n_\x^a$ (with x a label identifying the species), the functional dependence of the Lagrangian density\footnote{Obviously we here refer to the matter sector of the Lagrangian density.} becomes $\sqrt{|g|}\,\Lambda(n_\x^2,\,n_{\x\y}^2)$ where
\begin{equation}
\begin{split}
n_\x^2 &= - n_\x^a n_\x^b g_{ab} \;,\\
n_{\x\y}^2 &= - n_\x^a n_\y^b g_{ab} \;.
\end{split}
\end{equation}
An immediate feature of the variational approach is that it naturally provides the conjugate four-momenta of each species while accounting for the entrainment effect (see, for example, \cite{comer03:_rel_ent,Chamel:2017pql}). Roughly speaking, entrainment causes a species' four-momentum $\mu^\x_a = \partial\Lambda/\partial n^a_\x$ to be misaligned with its respective particle flux $n^a_\x$. We see this explicitly in the momentum/flux relation
\begin{equation}
	\mu^\x_a = \B^\x n^\x_a + \sum_{\y\neq\x} \A^{\x\y}n^\y_a \;,
	\label{mudef}
\end{equation}
where 
\begin{equation}
    	\B^\x = -2\frac{\partial\Lambda}{ \partial n_\x^2} \;,
\end{equation}
while the entrainment coefficients are defined as
\begin{equation}
    \A^{\x\y} = - \frac{\partial\Lambda}{ \partial n_{\x\y}^2} \;.
    \end{equation}
It is well established that, to obtain non-trivial (Euler-Lagrange) equations of motion from such a Lagrangian, the variation of the particle fluxes must be constrained \cite{SchutzSorkin1977,2004Prix}.

\subsection{The Matter Space Formulation and Field Equations}

A particularly elegant way of imposing the relevant constraint involves introducing the matter space, defined by identifying each current worldline as a single point, see \cite{CarterQuintana72}. For each fluid, the matter space is a three-dimensional manifold, so that when we introduce a set of coordinates $X_\x^A$ on, say, the $\x$-fluid's matter space, we give a ``name'', or label, to each fluid element. Because the entire worldline of each fluid element is mapped to a single matter space point,  it is clear that the fluid element's label $X^A_\x$, now considered a collection of three scalars on spacetime, takes the same value at each point on the worldline. After assigning a label to each fluid element worldline, we can use the linear map 
\begin{equation}\label{eq:MatterSpacelinearMap}
	\Psi^A_{\x\,a}\doteq \frac{\partial X_\x^A}{\partial x^a} 
\end{equation}
to push-forward (pull-back) vectors (co-vectors) between spacetime and the matter spaces. This is important because we can associate with each of the particle fluxes $n^a_\x$ a three-form $n^\x_{abc}$ by the standard Hodge-dual procedure\footnote{Here, we follow the Hodge-dual convention of \cite{GRCarroll} and have used
$\veps^{b c d a} \veps_{e b c d} = - 3! \delta^a_e $ to establish a sign convention for the dual $\veps^{b c d a}$ of the spacetime measure form.}: 
\begin{equation}
	n_\x^a = \frac{1}{3!} \veps^{bcda}\,n^\x_{bcd} \;,\quad n^\x_{abc} = \veps_{eabc}\,n_\x^e \;.
\end{equation}
 Now we can assume that the spacetime three-form $n^\x_{abc}$ is obtained by pulling back a corresponding matter space three-form, to be denoted $n^\x_{ABC}$; namely,
\begin{equation}
	n^\x_{abc} = \Psi^A_{\x\,[a}\Psi^B_{\x\,b}\Psi^C_{\x\,c]}n^\x_{ABC} \;,
\end{equation}
where, as usual, straight brackets indicate anti-symmetrization (and round ones symmetrization). Similarly, upon applying the Hodge-dual process to the four-momentum $\mu^\x_a$, we can push-forward with the map and identify a matter space momentum ``three-form'' $\mu^{ABC}_\x$ via
\begin{subequations}
\begin{align}
	\mu_\x^{abc} &=\veps^{dabc}\,\mu^\x_d	\;,\\
    \mu_\x^{ABC} &= \Psi^A_{\x\,[a}\Psi^B_{\x\,b}\Psi^C_{\x\,c]} \,\mu_\x^{abc}\;.
\end{align}
\end{subequations}
The main idea of the convective variational principle is to obtain the particle flux variation $\delta n_\x^a$ by first varying the matter-space three-form and then working backwards. 

Generally speaking, there are two ways of tracking changes in a fluid system---Eulerian and Lagrangian. The first, to be denoted by a $\delta$, measures changes in the fluid with respect to frames defined by the spacetime coordinates. The second, to be denoted $\Delta_\x$, measures changes with respect to fluid elements. Locally, the two can be related through the Lie derivative along some displacement vector field. If $\xi^a_\x$ is the displacement, then\footnote{We note that this relation between Lagrangian and Eulerian variation works to first order in $\xi^a_\x$, see \citet{Friedman1978a} for further details.}
\begin{equation}
\Delta_\x = \delta + \mathcal{L}_{\xi_\x} \; ,
\end{equation}
where $\mathcal{L}_{\xi_\x}$ is the Lie derivative with respect to $\xi^a_\x$. Because the label $X^A_\x$ of a fluid element is fixed, we can assert
\begin{equation}
	\Delta X^A_{\x} = 0 
	\; \Longrightarrow \; \delta X^A_{\x} = - \mathcal{L}_{\xi_\x} X^A_{\x} = - \Psi^A_{\x\,a} \xi^a_\x \; ,
\end{equation}
and thereby lock any displacement vector $\xi^a_\x$ on spacetime to a variation $\delta X^A_\x$ on matter space. Now, it is easy to show that the particle flux variation $\delta n^a_\x$ is (see \cite{2015CQAnderssonComer}) 
\begin{equation}\label{eq:ConvectiveVariation4Current}
	\delta n_\x^a = -\frac{1}{2}n_\x^a g^{bc}\delta g_{bc} -\frac{1}{3!} \veps^{bcda}\bigg(\mathcal{L}_{\xi_\x} n^\x_{bcd} - \Psi^B_{\x\,[b}\Psi^C_{\x\,c}\Psi^D_{\x\,d]}\,\Delta_{\x} n^\x_{BCD} \bigg) \;,
\end{equation}

Formally, we can take $n^\x_{ABC}$ to be a particle measure form on the matter space, which ``counts'' the total number of species $\x$ particles in the system. If it is a tensor on matter space then it must be  a function only of the matter space coordinates $X^A_\x$. But recall that in spacetime, the matter space labels are scalar fields $X^A_\x\left(x^a\right)$ with the special property that they take the same value at each spacetime point along the  fluid element worldline with which they are associated; clearly, this means that $n^\x_{ABC}$ also takes the same value with respect to its particular fluid element worldline. The net impact is that $n^\x_{abc}$ is automatically closed---because $n^\x_{ABC}$ is a three-form on a 3-dimensional matter space---and therefore the particle flux is conserved\footnote{Here with $dn$ we mean the exterior derivative of the differential form $n$.}:
\begin{equation}\label{eq:conservativeGamma}
\nabla_a n_\x^a = \frac{1}{3!}\veps^{bcda} \nabla_{[a}n^\x_{bcd]}  = \frac{1}{4!}\veps^{bcda} (dn)_{abcd} =  0 \;.
\end{equation}
But there is a deeper point to be made here.

The fact that $n^\x_{ABC} = n^\x_{ABC} (X^A_\x)$ implies $\Delta_\x n^\x_{ABC}= 0$, and one can verify that the flux variation above---as well as the equations of motions that follow---reduces to the well-known result for non-dissipative fluids (see \cite{NilsGregReview}). Therefore, to get the non-dissipative equations of motion one simply has to impose that the number of particles is conserved in the variation, or, equivalently, that the particle creation rates $\Gamma_\x = \nabla_a n^a_\x$ vanish. It then follows \cite{2015CQAnderssonComer} that a way to include dissipative processes (read: $\Gamma_\x \neq 0$) at the level of the action principle is to break the matter space tensorial nature of the particle measure form $n^\x_{ABC}$ and allow it to be a function of more than just the $X^A_{\x}$; in other words, we break the closure property of the $n^\x_{abc}$.

This general strategy has been applied by
\citet{2015CQAnderssonComer}, who focused on the cases where $n^\x_{ABC}$ depends on other species matter space coordinates $X^A_\y$ as well as the projected metrics
\begin{subequations}
\begin{align}
	g_\x^{AB} &= \Psi^A_{\x\,a} \Psi^B_{\x\,b}\,g^{ab} \;,\\
    g_\y^{AB} &= \Psi^A_{\y\,a} \Psi^B_{\y\,b}\,g^{ab} \;,\\
    g_{\x\y}^{AB} &= \Psi^A_{\x\,a} \Psi^B_{\y\,b}\,g^{ab} \;.
\end{align}	
\end{subequations}
This additional functional dependence in the particle measure forms $n^\x_{ABC}$ turns out to produce additional terms in the equations of motion representing different dissipation channels. Specifically, the equations of motion take the form
\begin{equation}\label{eq:DissipativeVariationalEoM}
	f^\x_a + \Gamma_\x \mu^\x_a = -\nabla^b D^\x_{b a} + R^\x_a \;,
\end{equation}
where
\begin{subequations}
\begin{align}
	f^\x_a &= 2 n_\x^b \nabla_{[b}\mu^\x_{a]} \;, \\
	D^\x_{ab} &= S^\x_{ab} + \sum_{\y\neq\x} s^{\y\x}_{ab} + \frac{1}{2}\Big(\mathcal{S}^{\x\y}_{ba} + \mathcal{S}^{\y\x}_{ab}\Big) \;, \\
    R^\x_a &= \sum_{\y\neq\x} \bigg(\text{R}^{\y\x}_a - \text{R}^{\x\y}_a\bigg) + \bigg(r^{\y\x}_a-r^{\x\y}_a\bigg) + \bigg(\mathcal{R}^{\y\x}_a -\mathcal{R}^{\x\y}_a \bigg) \;.
\end{align}
\end{subequations}
while the total stress-energy-momentum tensor is
\begin{equation}
 T_{a b} = \Psi  g_{a b} + \sum_{\x} \left(n^\x_a \mu^\x_b + D^\x_{ab}\right) \;.
\end{equation}
Projecting the field equation along $u^a_\x = n^a_\x/n_\x$, we see that
\begin{equation}\label{eq:gamx}
	\left(- u^a_\x \mu^\x_a\right) \Gamma_\x = u^a_\x \nabla^b D^\x_{b a} - u^a_\x R^\x_a \; .
\end{equation}
We note that, since the action-based model treats the entropy current in the same way as the other particle fluxes, this last formula becomes central when a specific model is considered as we need to make sure it is compatible with the second law of thermodynamics. It also follows, as an identity, that $\sum_\x R^\x_a = 0$, and because of this we have automatically $\nabla^a T_{a b} = 0$. Finally, it must be that $u^b_\x D^\x_{ab} = 0$.  We stress that the equations of motion above are obtained as Euler-Lagrange equations from the action, with the stress-energy-momentum tensor built in the standard way as a variation of the Lagrangian (density) with respect to the metric.

The explicit expressions for the $\mathcal{R}^{\y\x}_a$, $S^\x_{ab}$, etcetera terms will be discussed later, at a point where their application is more relevant. However, it is important to note at this point that the ``resistive terms'' $r^{\x\y}_a,\,\mathcal{R}^{\x\y}_a$ as well as the viscous tensors $s^{\x\y}_{ab} ,\,\mathcal{S}^{\x\y}_{ab}$ arise because we assume that $n^\x_{ABC}$ depends on $g_\y^{AB}$ and $g^{AB}_{\x\y}$, respectively. Also, it is easy to see that in general the x-species total viscous tensor $D^\x_{ab}$ is not necessarily symmetric because $\mathcal{S}^{\x\y}_{ab}$ is not. This property is, however, not inherited by the total viscous tensor of the system. We must have $D_{ab}= \sum_\x D^\x_{ab} = D_{ba}$.

\subsection{Matter Space Volume Forms}

All dissipative terms that enter the action-based equations are obtained by assuming that the fundamental current three-forms $n^\x_{abc}$ depend on an additional set of quantities which breaks their closure ($\nabla_{[a} n^\x_{bcd]} \neq 0$). We now want to explain how this can happen, but begin by introducing a bit of notation. 

We need to distinguish between the Levi-Civita symbol $\eta_{ABC}$ and a volume measure form $\varepsilon^\x_{ABC}$ on the matter space. The Levi-Civita symbol is defined as $\eta_{ABC} = [A\,B\,C]$ for any chosen set of coordinates (and thus is not a tensor but a tensor density) while the volume measure form $\veps^\x_{ABC}$ can be defined\footnote{This is tricky for a couple of reasons: It is well known from work on general relativistic elastic bodies \cite{Karlovini1} that this is not the only possible choice. Also, the projected metric $g^{AB}_\x$ is not ``fixed'' in the sense that the spacetime metric $g_{ab}$ changes, in a general curved spacetime, as a fluid element moves from point-to-point along its worldline.} by means of the push-forward of the metric: 
\begin{subequations}
\begin{align}
		&g_\x = \frac{1}{3!} \eta_{ABC}\eta_{DEF}\,g_\x^{AD}	g_\x^{BE}g_\x^{CF} =\text{det} (g_\x^{AB}) \\
        &\veps^\x_{ABC} = \sqrt{g^\x}\eta_{ABC} = \sqrt{g^\x}[A\,B\,C]
\end{align}
\end{subequations}
where $g^\x = (g_\x)^{-1}$ is the determinant of the inverse matrix $g^\x_{AB}$; i.e.~$g^\x_{AC} g_\x^{CB} = \delta^B_A$. 

This volume form provides a way to measure the volume of ``matter elements'', infinitesimal volumes in the matter space manifold. We can relate these quantities to the current and momentum three-forms 
\begin{subequations}
\begin{align}
	n^\x_{ABC} &= \N_\x\,\varepsilon^\x_{ABC} = \bar \N_\x \eta^\x_{ABC} \,,\\
    \mu_\x^{ABC} &= \M_\x\,\veps_\x^{ABC} = \bar \M_\x \,\eta_\x^{ABC}  \;.
\end{align}	
\end{subequations}
The point we want to make here is that the barred quantities look more like scalar densities on the x-matter space, while the non-barred ones look more like scalars. The relation between the two normalizations is simply
\begin{subequations}\label{eq:normalizationsRel}
\begin{align}
	\N_\x &= \sqrt{g_\x}\,\bar \N_\x \; , \\
	\M_\x &= \sqrt{g^\x}\,\bar \M_\x \;.
\end{align}
\end{subequations}
We can use this to expedite our use of the convective variational principle by focusing the additional functional dependence of $n^\x_{ABC}$ into 
\begin{equation}
    \N_\x = \N_\x(X_\x^A,\,X_\y^A,\,g_\x^{AB},\,g_\y^{AB},\,g_{\x\y}^{AB})\ . \label{eq:mspnorm}
\end{equation}  

To make contact with proper quantities measured in spacetime---that is, with the rest frame density and momentum for each fluid component---it is useful to introduce an appropriate tetrad $e^{\ha}_a$ for each species; an orthonormal basis whose timelike unit vector ${\boldsymbol e}_{\hat 0} = {\boldsymbol u}_\x$, so that $u_\x^\ha =({\boldsymbol e}_{\hat 0})^\ha =\delta_{\hat 0}^\ha = (1,0,0,0)^T$.  The components of the spacetime measure form in this tetrad basis are\footnote{Recall that, since $g_{ab}=e^\ha_a e^\hb_b \eta_{\ha\hb}$, the determinant of the tetrad $ e = \sqrt{|g|}$.}
\begin{equation}
	\veps^{\ha\hb\hc\hd} = \veps^{abcd}  e^\ha_a e^\hb_b  e^\hc_c e^\hd_d = \eta^{\ha\hb\hc\hd}
\end{equation}
where $\eta^{\ha\hb\hc\hd} = - [\ha\, \hb\,\hc\,\hd ]$ and we have omitted the chemical index. Now, since push-forward (and pull-back) is a linear map between vector spaces (the tangent space), it transforms as a linear map under coordinate changes, and we can write
\begin{equation}
	A^A = \frac{\partial X_\x^A}{\partial x^a} A^a = \Psi^A_{\x\,\ha}A^\ha
\end{equation}
where we have introduced the short-hand notation\footnote{Following \cite{GRCarroll} we denote the inverse matrix of the tetrad as $e^a_\ha$.}
\begin{equation}
	\Psi^A_{\x\,\ha}   \equiv \Psi^A_{\x\,a} \, e^a_\ha  = \frac{\partial X^A_\x}{\partial x^a}  \, e^a_\ha \;.
\end{equation}
Making use of the fact that $0 = u^\ha_\x \Psi^A_{\x\,\hat a} =\Psi^A_{\x\,\hat 0}$ we then get\footnote{The index $\hi$  runs over the $1,2,3$ components of the tetrad basis, and $\ha = \hat 0, \hi$.}
\begin{equation}
	g_\x^{AB} = \Psi^A_{\x\,\ha} \Psi^B_{\x\,\hb}\eta^{\ha\hb} \Longrightarrow g_\x = \text{det}\big(\Psi^A_{\x\,\hi}\big)^2 \;,
\end{equation}
which leads to\footnote{Note that, because of the standard convention we use $\eta^{\hat 0\hb\hc\hd} = -\varepsilon^{\hb\hc\hd}$ with $\hb,\hc,\hd = 1,2,3$.} 
\begin{equation}
\begin{split}
	\M_\x &= \frac{1}{3!}\, \mu_\x^{ABC}\,\veps_{ABC}^\x = \\
        &= \frac{1}{3!} \sqrt{g^\x}\eta_{ABC} \, \Psi^A_{\x\,\ha}\Psi^B_{\x\,\hb}\Psi^C_{\x\,\hc} \, \varepsilon^{\hat 0\ha\hb\hc } \mu^\x_{\hat 0} = \mu_\x \;,
\end{split}
\end{equation}
where we have used $ \mu_\x = -\mu^\x_a u^a_\x = -\mu_\x^{\hat 0}$. This result is important because it makes clear that only the (rest-frame) energy content of the four-momentum co-vector $\mu^\x_a$ is stored in the normalization of the matter space momentum three-form $\mu_\x^{ABC}$. Similarly, one can show that $\N_\x = n_\x$, in fact
\begin{equation}
\begin{split}
	n_\x &= -\frac{1}{3!}u^\x_a\,\veps^{bcda}\,n^\x_{bcd} = -\frac{1}{3!} u^\x_\ha\,\veps^{\hb\hc\hd\ha}\,\Psi^B_{\x\,[\hb}\Psi^C_{\x\,\hc} \Psi^D_{\x\,\hd]} \,\veps^\x_{BCD}\N_\x \\
    &=-\frac{1}{3!}u^\x_{\hat 0} \,\veps^{\hb\hc\hd\hat 0} \veps_{\hb\hc\hd}\,\N_\x = - u^\x_{\hat 0} \N_\x = \N_\x \;.
\end{split}
\end{equation}
These relations are not surprising. It is quite intuitive that the  non-barred quantities are related to spacetime (rest-frame) densities given that the three-forms $\veps^\x_{ABC}$ measure the volume of the matter space elements. 

We can also use the tetrad formalism to prove another result that will be needed later on; the intimate connection between a non-zero particle creation rate and an extended functional dependence of the current three-form. In fact, we have (see \cref{eq:conservativeGamma})
\begin{equation}
\begin{split}
    \Gamma_\x = \nabla_a n_\x^a = \frac{1}{3!} \veps^{bcda} \,\Psi^B_{\x\,[b}\Psi^C_{\x\,c}\Psi^D_{\x\,d}\nabla_{a]} n^\x_{BCD} \;,
\end{split}
\end{equation}
where we used $\nabla_{[a}\Psi^B_{\x\,b}\Psi^C_{\x\,c}\Psi^D_{\x\,d]}=0$. Introducing (again) a tetrad comoving with the x-species, and multiplying by $\mu_\x$ we have 
\begin{equation}\label{eq:NonConservativeGamma}
	\mu_\x\Gamma_\x = \frac{1}{3!} \mu_\x^{ABC} \,u_\x^a\nabla_a n^\x_{ABC} \equiv \frac{1}{3!} \mu_\x^{ABC} \frac{d n^\x_{ABC}}{d \tau_\x} \;,
\end{equation}
$\tau_\x$ being the proper time along the x-species worldline. As explained earlier, the right-hand-side of this equation vanishes identically if $n^\x_{ABC} = n^\x_{ABC} (X^A_\x)$, while it is in general non-zero if we assume the extended functional dependence given in \cref{eq:mspnorm}.

We can now use the introduced normalizations to slim the notation (with respect to that used in \cite{2015CQAnderssonComer}) for the various pieces of $R^\x_a$ and $D^\x_{ab}$ which were introduced but not defined above. For instance, the ``purely resistive'' term from \cite{2015CQAnderssonComer} becomes
\begin{equation}\label{eq:PurelyReactiveSlim}
	\text{R}^{\x\y}_a = \frac{1}{3!} \mu_\x^{ABC} \frac{\partial n^\x_{ABC}}{\partial X_\y^D}\,\Psi^D_{\y\,a} = \M_\x\frac{\partial\N_\x}{\partial X_\y^D} \,\Psi^D_{\y\,a}\equiv \text R^{\x\y}_D\,\Psi^D_{\y\,a} \;. 
\end{equation}
Similarly we can write 
\begin{subequations}\label{eq:AdditionalViscousTensorsSlim}
\begin{align}
\begin{split}
		s^{\x\y}_{ab} &= \frac{1}{3} \mu_\x^{ABC} \frac{\partial n^\x_{ABC}}{\partial g_\y^{DE}}\,\Psi^D_{\y\,a} \,\Psi^E_{\y\,b} = 2\M_\x \frac{\partial \N_\x}{\partial g_\y^{DE}}\,\Psi^D_{\y\,a} \,\Psi^E_{\y\,b}   \\ 
	&\equiv s^{\x\y}_{DE}  \,\Psi^D_{\y\,a} \,\Psi^E_{\y\,b} \;, 
\end{split} \\
\begin{split}
	\mathcal{S}^{\x\y}_{ab}& = \frac{1}{3} \mu_\x^{ABC} \frac{\partial n^\x_{ABC}}{\partial g_{\x\y}^{DE}} \,\Psi^D_{\x\,a} \,\Psi^E_{\y\,b} = 2 \M_\x \frac{\partial\N_\x}{\partial g_{\x\y}^{DE}}\Psi^D_{\x\,a} \,\Psi^E_{\y\,b} \\
    &\equiv \mathcal{S}^{\x\y}_{DE}\,\Psi^D_{\x\,a} \,\Psi^E_{\y\,b} \;,
\end{split}
\end{align}
\end{subequations}
where we have used the fact that the partial derivatives are taken, say, with respect to the metric $g^{AB}_\y$ keeping fixed $g^{AB}_\x$ and $g^{AB}_{\x\y}$. We will consider the validity of this assumption later. The remaining viscous stress tensor, $S_{ab}^\x$, leads to a slightly more involved expression, because of the presence of $g^\x$ in \cref{eq:normalizationsRel}. We have
\begin{equation}\label{eq:StdViscousTensorSlim}
\begin{split}
	S^\x_{ab} &= \frac{1}{3} \mu_\x^{ABC} \frac{\partial n^\x_{ABC}}{\partial g_\x^{DE}}\,\Psi^D_{\x\,a} \,\Psi^E_{\x\,b} = 2\Bigg( \frac{\M_\x}{\sqrt{g^\x}} \frac{\partial \big(\N_\x\sqrt{g^\x}\big)}{\partial g_\x^{DE}} \Bigg)\,\Psi^D_{\x\,a} \,\Psi^E_{\x\,b} =\\
    &=  2\Bigg(\M_\x \frac{\partial \N_\x}{\partial g_\x^{DE}} - \frac{1}{2}\N_\x\M_\x\,g^\x_{DE}\Bigg)\Psi^D_{\x\,a} \,\Psi^E_{\x\,b} =\\
    &\equiv S^{\x}_{DE}  \,\Psi^D_{\x\,a} \,\Psi^E_{\x\,b}  \;.
\end{split}
\end{equation}
It is also obvious, by looking at the respective definitions, that the resistive terms that stem from the fact that $\N_\x$ can depend also on $g^{AB}_\y$ and $g^{AB}_{\x\y}$ can be now written  
\begin{subequations}
\begin{align}
	&r_a^{\x\y} = \frac{1}{2}s^{\x\y}_{DE}\, \nabla_a \big(g^{bc} \Psi^D_{\y\,b} \Psi^E_{\y\,c}\big) \;,\\
    & \mathcal{R}_a^{\x\y} = \frac{1}{2}\mathcal{S}^{\x\y}_{DE}\, g^{bc} \Psi^D_{\x\,b} \nabla_a \big( \Psi^E_{\y\,c}\big) \;.
\end{align}
\end{subequations}

Before moving on, it is useful to consider the simplest non-dissipative fluid model that can be derived from the action above---the ordinary perfect fluid, where all particle species and entropy flow together and the total particle numbers and entropy are conserved individually. The calculation is straightforward \cite{comer93:_hamil_multi_con}. All fluxes have the same four-velocity, say, $u^a$, and so $n^a_\x = n_\x u^a$. If each particle number flux is conserved individually, then
\begin{equation}
    \nabla_a n^a_\x = \nabla_a \left(n_\x u^a\right) 
      = u^a \nabla_a n_\x + n_\x \nabla _a u^a = 0 
      \; \Longrightarrow u^a \nabla_a \ln n_\x = - \nabla _a u^a \ .
\end{equation}
Obviously, the total particle flux $n^a = \sum_\x n^a_\x$ is also conserved and so we can write as well
\begin{equation}
     u^a \nabla_a \ln n = - \nabla _a u^a \; , \; n = \sum_\x n_\x \ .
\end{equation}
Therefore, we have
\begin{equation}
     u^a \nabla_a \ln n_\x - u^a \nabla _a \ln n = 0 
      \; \Longrightarrow u^a \nabla_a \Big(\frac{n_\x}{n}\Big) = 0 \ .
\end{equation}
The upshot is that each species fraction $n_\x/n$ must also be conserved along the flow, and this includes the entropy as well. This implies that only one matter space is required. In the action principle, this means that for each $\x$ we have $\xi^a_\x = \xi^a$, and there is only one Euler equation of the form
\begin{equation}
     \sum_\x f^\x_a = 0 \ , \label{singfl}
\end{equation}
where the $f^\x_a$ are exactly as defined before\footnote{The general case of a two-component model is discussed in \cite{NilsGregReview}.}.

\section{The Non-dissipative Limit}\label{sec:thermoEq}

We now begin to develop the process for comparing standard relativistic models for dissipative fluids with that provided by the action principle. Standard approaches \cite{Muller:1967zza,ISRAEL1976,transientStewart77,IsraelStewart79,IsraelStewart79bis} start with a definition of equilibrium and then build in dissipation via deviations away from this state. Conversely, the action principle provides a fully non-linear set of field equations, while it does not require any sort of equilibrium. Obviously, our first task must be to extract from the non-linear equations a notion of equilibrium. This is not straightforward as an arbitrary spacetime in General Relativity does not have global temporal, spatial, and rotational invariance.

\subsection{General Relativistic Set-up}
A typical laboratory set-up is local in the spacetime sense, as long as the effect of long-range, non-screenable forces on the system---for example, gravity---can be ignored. Well-defined (theoretical and experimental/observational) notions of total energy and entropy can be realized. Equilibrium can be defined in the broadest sense by saying that the system evolves to a state where its total energy is minimized, or, equally, its total entropy is maximized.
In contrast, a general relativistic set-up is problematic from the get-go, because one is hard-pressed to find properties of equilibrium which are workable at all time- and length-scales. Broadly speaking, there seems to be no general relativistic rules on how the local thermodynamics of local (intensive) parameters---chemical potential $\mu$, pressure $p$, and temperature $T$---connects with some notion of global thermodynamics for global (extensive) parameters---such as the total energy $E$. An unambiguous extrapolation of the standard definitions of chemical,  dynamical, and thermal equilibrium to General Relativity is not possible, for reasons to be explained below. There is also the well-known difficulty of identifying the total energy of a region in an arbitrary spacetime, since the Equivalence Principle precludes an ultra-local definition of gravitational energy density.\footnote{Of course, for asymptotically flat spacetimes, one can define quantities like the Schwarzschild mass. Gravitational-wave's energy can be defined but only after averaging over wavelengths.}

The reason that the laboratory rules for chemical and thermal equilibrium are not viable in General Relativity was established long ago by Tolman and Ehrenfest \cite{TOLMAN1,TOLMAN2}: In General Relativity, all forms of energy react to gravity. Temperature and chemical potentials represent forms of energy and can undergo red-shift or blue-shift. There is no one temperature for an isolated system, and so statements like ``system A is in thermal equilibrium with system B if their temperatures are the same'' become ambiguous; similarly for chemical equilibrium. 

Even the use of the word ``equilibrium'' becomes problematic because it tends to imply that a system in thermal and chemical equilibrium is independent of time, because the total entropy and total particle number do not evolve. In General Relativity, systems which are independent of time corresponds to spacetimes with a global timelike Killing vector field.
Strictly speaking, this immediately puts the non-dissipative fluid models of Cosmology---the Friedman-Lemaitre-Robertson-Walker solution---out of the discussion, as the universe is expanding, making it time-dependent.

The main message is this: Important issues remain unsettled even after a century's worth of debate. We will not resolve these issues here; instead, what we will do is take the action-based formalism and see how its internal machinery can be manipulated to produce a self-consistent notion of the non-dissipative limit, without trying to resolve deeper issues about the nature of equilibrium.\footnote{We will still use the word ``equilibrium'' interchangeably with the non-dissipative limit.} Our way forward is to take advantage of the fact that the action-based field equations are fully non-linear and complete. 

\subsection{Multiple Equilibrium States}

The main mechanism for manipulating the machinery of the action-based field equations is to apply perturbation techniques similar to those used to determine, say, quasi-normal modes of neutron stars. The general idea for neutron stars is to analyze linear perturbations of configurations having particular symmetries generated by Killing vectors. Among the most studied neutron star ``ground-states'' are those having Killing vectors which generate staticity and spherical symmetry, and those with Killing vectors that generate axisymmetry and stationarity; basically, non-rotating and rotating backgrounds, respectively.

In an analogous way, we can expect different options for generating the non-dissipative limit of a multi-fluid system. For example, we can take the limit where the different dissipation coefficients (such as shear and bulk viscosities) are effectively zero. Another possibility is the limit where the dissipation coefficients are non-zero but the fluid motion itself is such that the dissipation mechanisms are not acting. The formalism developed by \citet{onsager31:_symmetry} is worthy of mention here, because the system of field equations it creates are more explicit in how the two limits can be implemented (see, for example, \cite{andersson05:_flux_con}). It is also interesting to note that the philosophy of the Onsager approach is not so much about how to expand away from an equilibrium, but rather how a non-equilibrium system gets driven back to the equilibrium state. Here, because the field equations are fully non-linear, they can, in principle, describe systems which are being driven towards or away from equilibrium. For this reason, in the final model---i.e. when the various dissipative terms are fully specified---one has to insist on consistency with the second law of thermodynamics to ensure the system is evolving towards (and not away from) equilibrium.

Next, we will use a global analysis which assumes that the second law of thermodynamics applies and that a knowledge of the fluxes  throughout a region of spacetime is enough to determine whether or not dissipation is acting. A local analysis of the formalism will also be pursued, involving the field equations themselves. 
 
\subsection{Global Analysis of the Non-dissipative Limit}
Let us first of all recall the logic behind the fluid modelling scheme. The crux of it is to assume that knowledge of the total mass-energy and momentum flux obtained by tracking the worldlines of individual particles can be replaced with tracking the worldlines of fluid elements, which are defined in the following way: Fill-up side-to-side, top-to-bottom, and front-to-back the entire system with $I = 1 ... M$ local conceptual boxes---the fluid elements. Each element has its own volume $\delta V_I$, number of particles $\delta N_I$, mass-energy $\delta E_I$, and entropy $\delta S_I$. Clearly, as the number $M$ is increased the elements become ultra-local, implying that the change in the spacetime metric across them is small. 

Now consider the $I^{\rm th}$-fluid element. It moves through spacetime and, if the element is small enough, the trajectory  can be accurately represented by a single unit four-velocity $u^a_I$. When taken together, and in the limit $M \to \infty$, all the $u^a_I$ form a vector field on spacetime and this field plays a role in the fluid system's degrees of freedom. If a local typical scattering length $\lambda_I$ between the particles exists, and the size of fluid elements is commensurate with that length ($\delta V_I \sim \lambda^3_I$), then the average four-velocity of the $\delta N_I$ particles will be $u^a_I$. 

In principle, we now have everything we need to define the actual fluid degrees of freedom, which are the particle fluxes $n^a_I = (\delta N_I/\delta V_I) u^a_I$. But, the fact that we have introduced typical scattering lengths and average velocities as part of our fluid element definition means we have assumed that fluid elements contain enough particles to warrant a statistical/thermodynamical treatment; i.e., we have to know how the individual four-momenta of the particles are distributed initially with respect to the fluid elements, and then redistributed as the fluid evolves. It is worth noting that there is nothing in this construction that limits the validity of the fluid modelling scheme to a close-to-thermodynamic-equilibrium regime.

Let us now consider the impact of equilibrium on fluid elements' space- and time scales. Here, because we impose the second law of thermodynamics below, we specifically identify the entropy flux $s^a$ in this discussion. The formalism's linchpin is the breaking of the closure of the particle-flux three-forms, $n^\x_{abc}$ and $s_{abc}$, which leads to non-zero creation rates $\Gamma_\x$ and $\Gamma_\s$. In turn, these non-zero creation rates lead to the resistive contribution, $R^a_\x$, and the dissipation tensor, $D^\x_{ab}$, terms in the equations of motion. 

\begin{figure}
   \centering
   \includegraphics[width=0.7\linewidth]{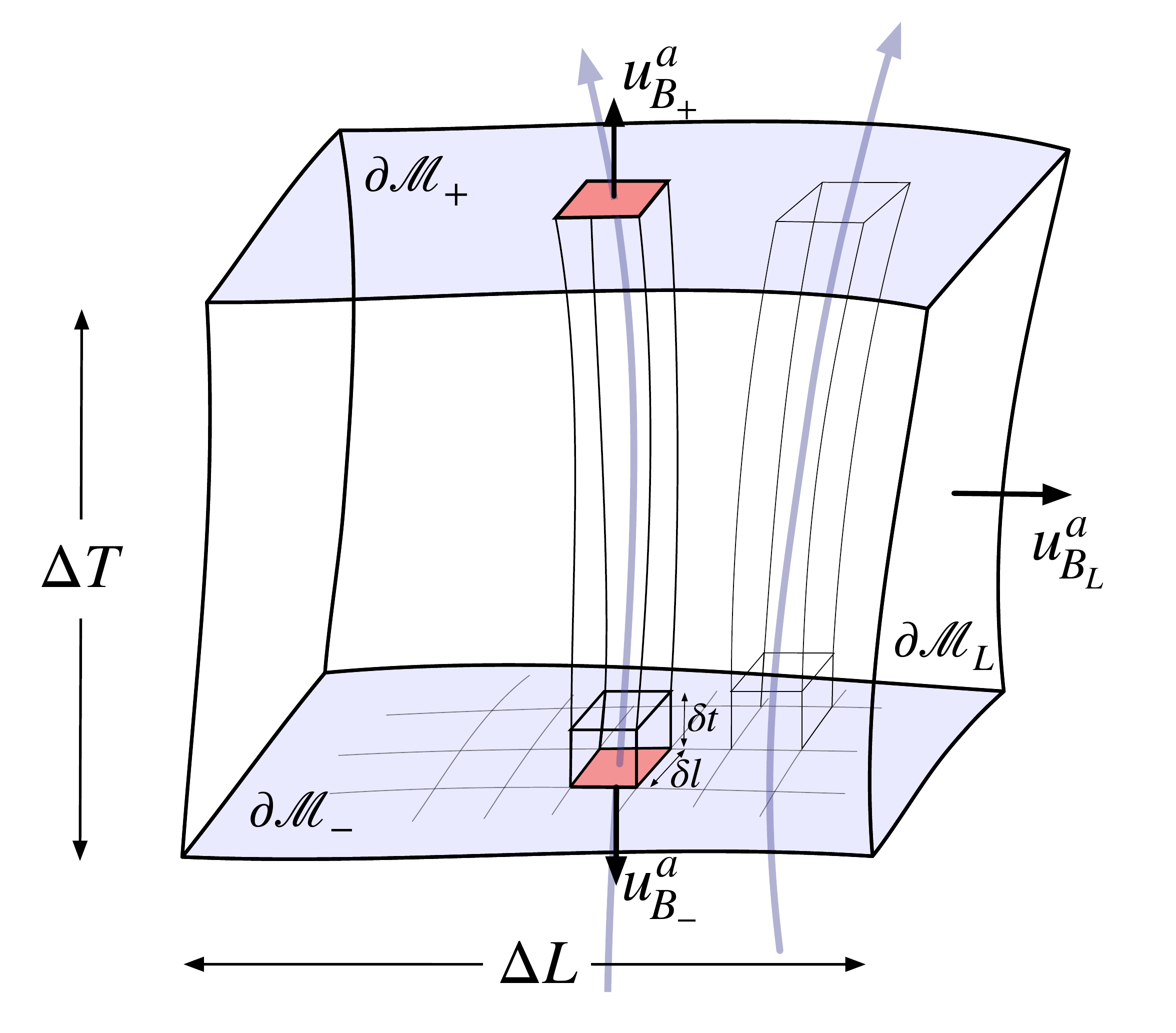}
   \caption{A depiction of the spacetime region ${\cal M}$, with one spatial axis suppressed. The region has a characteristic spatial size $\Delta L$ and temporal size $\Delta T$. Inside ${\cal M}$ is a smaller region $\delta {\cal M}$ of characteristic spatial and temporal size $\delta l$ and $\delta t$, respectively. The boundary $\partial\cal M$ consists of the initial and final time-slices $\partial\cal M_-,\,\partial\cal M_+$ and the timelike hypersurface $\partial\mathcal{M}_L$. }
   \label{fig:FluidElem}
\end{figure}

When we use the Einstein equations and the field equations of a multi-fluid system, our goal is to get solutions for the metric and fluxes on a ``chunk'' of spacetime, for a given set of initial/boundary conditions. Suppose we pick an ad hoc region ${\cal M}$ of spacetime, as illustrated in \cref{fig:FluidElem}. The fact that it is a specified region implies there is a ``conceptual boundary'', meaning the whole spacetime is being divided up into smaller domains. Let $u^a_{\rm B}$ (collectively) denote the unit normal to the total boundary of the region, defined so that it always points ``out''. The boundary itself consists of two spacelike hypersurfaces $\partial {\cal M}_\pm$ (with unit normals $u^a_{\rm B_\pm}$, $u^a_{\rm B_\pm} u_a^{\rm B_\pm} = -1$), and a timelike hypersurface $\partial {\cal M}_L$ (with unit normal $u^a_{\rm B_L}$, $u^a_{\rm B_L} u_a^{\rm B_L} = + 1$); in essence, think of $\partial {\cal M}_-$ as a 3D region of characteristic volume $\Delta L^3$ on an initial time-slice of ${\cal M}$ and $\partial {\cal M}_+$ as the same volume on the final time-slice, and then $\partial {\cal M}_L$ will be similar to the union of the surface of the same volume on each leaf of some spacelike foliation of ${\cal M}$ between $\partial {\cal M}_-$ and $\partial {\cal M}_{+}$. The induced metric on $\partial {\cal M}_{\pm}$ is $h_\pm^{a b} = g^{ab} + u^a_{\rm B_\pm} u^b_{\rm B_\pm}$ and for $\partial {\cal M}_L$ it is $h_L^{a b} = g^{ab} - u^a_{\rm B_L} u^b_{\rm B_L}$.

There are three contributions to the total particle number change $\Delta N^\x$\footnote{We have taken into account the fact that $u^a_{\rm B_-}$ points to the past.}
\begin{subequations} \label{fluxdelnx}
\begin{align}
	N_+^\x &= \tintp n^\x_+ \sqrt{h_+}\,\text{d}^3x= \tintp \left(- u_a^{\rm B_+} n^a_\x\right)\sqrt{h_+}\,\text{d}^3x \;,\\ 
	N_-^\x &= \tintm n^\x_- \sqrt{h_-}\,\text{d}^3x = \tintm \left(u_a^{\rm B_-} n^a_\x\right) \sqrt{h_-}\,\text{d}^3x \;, \\
	\Delta N_L^\x &= \tintL n^\x_L \sqrt{h_L}\,\text{d}^3x= \tintL \left(u_a^{\rm B_L} n^a_\x\right) \sqrt{h_L}\,\text{d}^3x\;,
\end{align}
\end{subequations}
and similarly for $\Delta S$.

The changes in the total $\x$-particles $\Delta N^\x$ and entropy $\Delta S$ over the region ${\cal M}$ are therefore
\begin{subequations} \label{totdelNx}
\begin{align}
\Delta N^\x = N_+^\x - N_-^\x + \Delta N_L^\x \ , \\
\Delta S = S_+ - S_- + \Delta S_L \ .
\end{align}
\end{subequations}

If the length- and time-scales of spacetime region ${\cal M}$ are those typical of terrestrial laboratories (read: its curvature is zero throughout), then we have confidence in asserting the second law of thermodynamics; namely, the net change of the total entropy must satisfy $\Delta S \geq 0$. We could even be confident that we can determine the total energy $E$ and volume $V$ of the system, and have a working First Law of Thermodynamics which connects $\Delta E$, $\Delta N^\x$, $\Delta V$, and $\Delta S$: 
\begin{equation}
    \Delta E = T \Delta S - p \Delta V + \sum_\x \mu_\x \Delta N^\x \ . \label{1stlaw}
\end{equation}
The temperature $T$, pressure $p$, and chemical potentials $\mu_\x$ would be well-defined and calculable. We could even use the standard notions of chemical, dynamical, and thermal equilibrium and say that system A of spacetime region ${\cal M}_A$ is in chemical, dynamical, and thermal equilibrium with system B of spacetime region ${\cal M}_B$ if, respectively, their chemical potentials are equal, their pressures are equal, and their temperatures are equal.  

Now, let us suppose we have a region large enough that spacetime curvature can no longer be ignored. Probably, it would be a safe bet to say that the second law still applies; i.e., $\Delta S \geq 0$. But, we are hard-pressed to employ the laboratory definitions of chemical, dynamical, and thermal equilibrium. Consequently, it is difficult to imagine a global First Law of Thermodynamics for general relativistic multifluid systems similar to that in \cref{1stlaw}; again, the reason being that intensive parameters are spacetime dependent, and an extensive parameter like total energy may not even be definable. Still, our task is to explore any possible link between parameters which require scales where spacetime curvature is necessary ($\Delta N^\x$ and $\Delta S$) to the local fluid variables ($n^a_\x$ and $s^a$) which enter the fluid field equations. Fortunately, the divergence theorem provides such a link.
Applying it to the divergence of both the particle and entropy fluxes we find
\begin{subequations}
	\begin{align}
		\Delta N^\x &= \fint \nabla_a n^a_\x\, \sqrt{-g}\,\text{d}^4x = \fint \Gamma_\x\,\sqrt{-g}\,\text{d}^4x \;, \\
		\Delta S &= \fint \nabla_a s^a\, \sqrt{-g}\,\text{d}^4x = \fint \Gamma_\s\, \sqrt{-g}\,\text{d}^4x\; .
	\end{align}
\end{subequations}
These are not new results, but they serve the purpose of establishing a direct link between global and local variables, which we will use to formulate some aspects of the non-dissipative limit of our formalism.

Consider an idealized situation of a spacetime region ${\cal M}$ such that sub-divided into a region ${\cal M}_A$ for which $\Delta N^\x_A < 0$ and $\Delta S_A < 0$, and another region ${\cal M}_B$ for which $\Delta N^\x_B > 0$ and $\Delta S_B > 0$, and assume they are such that the total changes on ${\cal M}$ vanish:
\begin{equation}
       \Delta N^\x = \Delta N^\x_A + \Delta N^\x_B = 0
       \; , \;
       \Delta S = \Delta S_A + \Delta S_B = 0 \ .
\end{equation}
The point is that, even though $\Gamma_\x$ and $\Gamma_\s$ are not zero in this case, this is an example of a global, fully general relativistic, non-dissipative system since there is no net total particle number or total entropy change. 

However, this is not the kind of limit we are aiming for. The non-dissipative limit is better understood by breaking up ${\cal M}$ into small spacetime regions $\delta {\cal M}$, with characteristic temporal and volume scales $\delta t$ and $\left(\delta l\right)^3$, respectively, as illustrated in \cref{fig:FluidElem}.
Once again, let us imagine that $\delta {\cal M}$  is subdivided into two regions $\delta {\cal M}_A$ and $\delta {\cal M}_B$. It is conceivable that on these scales statistical fluctuations could lead to positive creation rates in one region and negative in the other. If the regions are small enough, we can assume that $\Gamma_\x$ and $\Gamma_\s$ vary slowly across them so that we can approximate the integrals for $\delta N^\x_{\delta {\cal M}}$ and $\delta S_{\delta {\cal M}}$ as
\begin{equation}
       \delta N^\x_{\delta {\cal M}} \approx \Gamma_\x \delta t \left(\delta l\right)^3 
       \; , \;
       \delta S_{\delta {\cal M}} \approx \Gamma_\s \delta t \left(\delta l\right)^3 \ .
\end{equation}
However, the random nature of statistical fluctuations for a system purported to be in equilibrium implies that any non-zero creation rates inside $\delta {\cal M}_A$ and $\delta {\cal M}_B$ must balance on average so that
\begin{subequations}
       \begin{align}
       \delta N^\x_{\delta {\cal M}} = \delta N^\x_{\delta {\cal M}_A} + \delta N^\x_{\delta {\cal M}_B} \approx \left(\Gamma^A_\x + \Gamma^B_\x\right) \delta t \left(\delta l\right)^3 = 0 
       \; \Longrightarrow \; \Gamma_\x = \Gamma^A_\x + \Gamma^B_\x = 0 \; , \\
       \delta S_{\delta {\cal M}} = \delta S_{\delta {\cal M}_A} + \delta S_{\delta {\cal M}_B} \approx \left(\Gamma^A_\s + \Gamma^B_\s\right) \delta t \left(\delta l\right)^3 = 0 
       \; \Longrightarrow \; \Gamma_\s = \Gamma^A_\s + \Gamma^B_\s = 0 \; \ .
       \end{align}
\end{subequations}

One conclusion from this exercise is that the characteristic time and space scales of $\delta {\cal M}$ must be large enough  that statistical fluctuations will, on average, balance out for a system in equilibrium. The second conclusion is that having $\delta N^\x_{\delta {\cal M}} = 0$ ($\delta S_{\delta {\cal M}} = 0$) on the one hand means $\Gamma_\x = 0$ ($\Gamma_\s = 0$) on the other, and vice versa. Putting both together we will assume that the equilibrium state for multi-fluid systems must be such that regions like $\delta {\cal M}$ set the scales for fluid elements and $\Gamma_\x = 0$ and $\Gamma_\s = 0$ everywhere in ${\cal M}$.

\subsection{Local Analysis of the Non-dissipative Limit}
The next step begins where the previous one left off; that is, a necessary condition for a multi-fluid system to be in equilibrium is that the flux creation rates $\Gamma_\x$ (now including the entropy) vanish everywhere. We will use the field equations themselves to investigate the zero-dissipation limit further. But before we do this, we will impose another condition which defines the equilibrium. We require all distinct fluids to be comoving---e.g. we are not considering systems with superfluid/superconducting phases, or a perfect heat-conducting limit \cite{CarterNoto}. This means that there is a common four-velocity for all species. This is quite intuitive, but it is important to point out a subtlety about this comoving limit: For a multi-fluid system each species has its own evolution equation. Even in the comoving limit there are still $\x$ fluid equations. Now consider the field equations for a multi-species, single fluid system---as we see in \cref{singfl}, it has only one fluid evolution equation. Therefore, the comoving limit of the multi-fluid system ($\x$ equations) is not equal to the single-fluid system (one equation). This is not an error, rather, it is a consequence of the fact that the number of independent field equations of the system is fixed by the number of independent fluids chosen before the action principle is applied. Note also that we can use the common four-velocity $u^a$ to introduce a spatial covariant derivative $D_a$---acting in directions perpendicular to $u^a$---and a time derivative $``\;\dot{}\;" = u^a \nabla_a$. For a scalar $A$ we have
\begin{equation}\label{eq:ProjectedDerivative}
	D_a A = \perp^b_a \nabla_b A = \left(\delta^b_a + u_a u^b\right) \nabla_b A = \nabla_a A + \dot A u_a \; ,
\end{equation}
and for a vector
\begin{equation}\label{eq:ProjectedDerivativevec}
	D_a A_b = \perp^c_a \perp^d_b \nabla_c A_d \; ,
\end{equation}
where $\perp^b_a = \delta^b_a + u_a u^b$.

\subsubsection{Dynamical Suppression of Dissipation}
We start by considering the consequences of the non-dissipative limit  if the fluid flow is such that the dissipation mechanisms are not triggered. If we look at each species creation rate we have
\begin{equation}
	\mu_\x \Gamma_\x = - R^\x_a \,u^a - D^\x_{a b} \nabla^a u^b = 0
\end{equation}
so that, summing over all species
\begin{equation}\label{eq:EquilibriumRigidMotion}
	\sum_\x \mu_\x \Gamma_\x = - \left(\sum_\x R^\x_a\right) \,u^a - D_{a b} \nabla^a u^b = - D_{ab} D^{(a} u^{b)} = 0 \; ,
\end{equation}
where we have used the identities $u^b_\x D^\x_{ab} = 0$, $\sum_\x R^\x_a = 0$. Using the standard decomposition
\begin{subequations}
\begin{align}
    \nabla_a u_b &= -\dot u_b u_a + \varpi_{ab} + \sigma_{ab} + \frac{1}{3}\theta \perp_{ab} \; , \\
    \varpi_{ab} &= \perp^c_{[a} \perp^d_{b]} \nabla_c u_d \; , \\
    \sigma_{ab} &= \perp^c_{(a} \perp^d_{b)} \nabla_c u_d - \frac{1}{3} \theta \perp_{a b} \; , \\
    \theta &= \nabla_a u^a = D_a u^a \; ,
\end{align}
\end{subequations}
it is easy to see that \cref{eq:EquilibriumRigidMotion} implies 
\begin{equation}
   D_{(a} u_{b)} = \perp^c_{(a} \perp^d_{b)} \nabla_c u_d = \nabla_{(a} u_{b)} + u_{(a} \dot{u}_{b)} = \sigma_{ab} + \frac{1}{3}\theta \perp_{ab} = 0 \; .
\end{equation}
In particular, this tells us that the (dynamically-suppressed) non-dissipative flow has zero expansion $\theta = 0$, and zero shear $\sigma_{a b} = 0$. What is left of the motion is captured by
\begin{equation}
    \nabla_a u_b = \varpi_{ab} - \dot{u}_b u_a \; \ ,
\end{equation}
which is consistent with rigid rotation.

From the definition of creation rates, we can now write
\begin{equation}
	\Gamma_\x = \nabla_a n_\x^a = \dot n_\x + n_\x \theta = \dot n_\x = 0 \;.
\end{equation}
Assuming a thermodynamical relation in the standard way, namely that the energy functional of the system is $\varepsilon = \varepsilon (n_\x)$, we see that the chemical potential of each species is $\mu_\x = \mu_\x(n_\x)$ and likewise for the pressure $p$. Therefore, we have $\dot \mu_\x = 0$ and $\dot p = 0$, as well. The proposed scenario is consistent with the minimum requirements for the system being non-dissipative---as explained above. This is not, however, the situation we will use as basis for the expansion.

\subsubsection{The Euler limit}\label{vanRDeq}
Later (in \cref{sec:EnergyMinimized}) we will use thermodynamics arguments to show that the dissipative terms all vanish at equilibrium: $D_{ab}^{\x,\ \e} \text{ and } R_a^{\x,\ \e}=0$---where we have introduced the superscript ``e''  to stress that the dissipative terms are evaluated at equilibrium, consistently with the notation used later on. We now consider the non-dissipative limit with these additional constraints and show its compatibility with the Euler equations. Since the fluids are comoving at equilibrium we have for the fluxes $n_\x^a = n_\x u^a_\e$, and so the four-momenta become
\begin{equation}
	\mu_a^\x = \Big(\mathcal{B}_\x n_\x + \sum_{\y\neq\x}\mathcal{A}_{\x\y}n_\y \Big) u^\e_a = \mu_\x u^\e_a \;.
\end{equation}
and the equation of motion for the x-species is
\begin{equation}\label{eq:AlignedForce}
	f^\x_a =2n_\x^b \nabla_{[b}\mu^\x_{a]}= n_\x\mu^\x \dot u^\e_a + n_\x\Big(u^b_\e u^\e_a + \delta^b_a\Big)\nabla_b\mu^\x = n_\x \mu^\x \dot u^\e_a + n_\x D_a\mu^\x = 0 \;.
\end{equation}
The first term in $f^\x_a$ then looks like the mass/energy per volume times the acceleration while we can show that the second one is a ``pressure-like'' term in the sense of being the gradient of a thermodynamic scalar. In fact, we have
\begin{equation} \label{muxcomov}
	\frac{\partial\Lambda}{\partial n_\x} = - \Bigg(\mathcal{B}_\x n_\x -\sum_{\y\neq\x} \mathcal{A}_{\x\y}n_\y^a u^\x_a \Bigg) =- \mu_\x
\end{equation}
and the sum of these terms provides the derivative of the total pressure $\Psi$:
\begin{equation}
	\sum_\x n_\x D_a \mu^\x  = D_a \Big( \sum_\x n_\x \mu^\x +\Lambda\Big) = D_a \Psi \;.
\end{equation}
It is important to note that each individual term cannot (in general) be considered as the derivative of the x-species contribution to the total pressure. Partial pressures exist only when the various species do not interact.

Even though the comoving limit of the multi-fluid system is not the same as the single fluid, multi-species system, there is some overlap: Taking the sum over the chemical species of \cref{eq:AlignedForce} we find\footnote{We have used the standard Euler relation  $\sum_\x n_\x\mu^\x = p +\varepsilon$, where $p,\,\varepsilon$ are the equilibrium pressure and energy density, respectively.}
\begin{equation}
	(p + \varepsilon) \dot u^\e_a = - D_a p \;. \label{comoveeuler}
\end{equation}
This is the standard Euler form. One can show also that \cref{singfl} can be written in this form. This is an important self-consistency check, but because the multi-fluid comoving limit is not the same as the single fluid limit, we need to go back to the individual fluid equations of the multi-fluid system.

We can rewrite the individual equations of motion as
\begin{equation}\label{eq:MultiFluidEquilibriumEq}
	\dot u^\e_b = - D_b (\text{log }\mu_\x) \ ;
\end{equation}
thus, for each combination of $\x \neq \y$,
\begin{equation}
    D_a (\text{log }\mu_\x) = D_a (\text{log }\mu_\y) \; \Longrightarrow \; D_a \left(\text{log }\frac{\mu_\x}{\mu_\y}\right) = 0 \ .
\end{equation}
This self-consistency therefore requires the various chemical potentials $\mu_\x$ and $\mu_\y$ (as functions on spacetime) to be proportional to each other by some factor $C^\x_\y$, which is constant in the spatial directions; namely,
\begin{equation}
    \mu_\x = C^\x_\y \mu_\y \; , \; D_a C^\x_\y = 0 \ .
\end{equation}
This is to be contrasted with the single-fluid case, where there is no such restriction---in the sense of being forced by the evolution equations---between the chemical potentials. Usually, one must provide additional information. For example, for neutron stars one typically imposes that beta decay and inverse beta decay are in equilibrium. 
If we combine this with the ``dynamical suppression of dissipation'', the self-consistency requires the chemical potentials to be proportional to each other by some factor which is constant in all the space-time directions. 

\subsubsection{Dynamical Suppression and Killing Vectors}

In a local region of spacetime, freely falling frames exist and the Killing equation will be satisfied approximately. In these local regions having an equilibrium will be consistent with the existence of Killing fields. However, local regions which are far removed from each other will not be (on the relevant dynamical timescale) in equilibrium with each other. This kind of ``quasi-local'' regression towards equilibrium has been discussed in the work of \citet{FukumaSakataniEntropic}, explicitly introducing two different spacetime scales to describe the  evolution of general relativistic dissipative systems. The hypothesis of Local Thermodynamic Equilibrium applies on the smaller scale---which is of the size of the fluid element---while the regression (in the sense of Onsager \cite{onsager31:_symmetry}) towards equilibrium takes place on the larger one, which can still be smaller than the body size.

A relation between the perfect fluid four-velocity and Killing vectors, for stationary axially symmetric rotating stars,\footnote{Note that \citet{RotatingStarsGourgoulhon} works with the enthalpy per particle instead of chemical potentials. However, this makes no difference for barotropic perfect fluids.} has been discussed by \citet{RotatingStarsGourgoulhon}. A similar discussion about thermodynamic equilibrium in General Relativity and the existence of Killing vectors was provided by \citet{Becattini16}. Specifically, he showed that there must be global Killing vector fields if the total entropy of the system is to be independent of the spacelike hypersurface over which the integration is performed. As for the work presented here, we will now show under what conditions the combination $\xi^a_\x = \mu_\x^{-1} u^a_\e$ can be turned into Killing vector fields. 

From \cref{eq:MultiFluidEquilibriumEq} it can be seen that 
\begin{equation}
\begin{split}
	\nabla_a \xi_b^\x + \nabla_b \xi_a^\x = \frac{1}{\mu_\x} \big(u^\e_{(a}u^\e_{b)} \dot{\text{log} \mu_\x} + D_{(a}u^\e_{b)}\big)
\end{split} 
\end{equation}
so that, if dynamical suppression has worked, the right hand side becomes zero and the $\xi^a_\x$ will be a timelike Killing vector field, along which the local thermodynamical parameters $n_\x$, $\mu^\x$, $\veps$, and $p$ become constants of motion. Put in different words, if we want the system to be (at least quasi-locally) at equilibrium---i.e. stationary--- we also need to require rigid body motion.

\subsection{A Final Comment on Equilibrium}

To conlcude we will come full circle and consider again the change in total entropy given by \cref{totdelNx}. The result only references spacelike hypersurfaces as part of the (ad hoc) choice of the boundary of the spacetime region for which the entropy change is being determined. There are no restrictions placed on the spacetime geometry in this construct; in particular, no requirement of global Killing vectors. 

As a matter of practice, the change in entropy of a system is clearly dependent on the spatial size and the amount of time the system has had to evolve. Coupling this with the fact that a separation of space from time is always a choice---an arbitrary spacetime has no preferred directions, no natural ``moments-of-time"---we see that the ad hoc nature of the boundary in \cref{totdelNx} is not a drawback. This is precisely the freedom needed in order to incorporate a system's spatial extent and evolution time, and the fact that a separation of space from time in spacetime is always a choice.

The main reason why this is intriguing is that the second law of thermodynamics only refers to the change in total entropy, not the value of entropy itself at specific moments of time (i.e.~spacelike hypersurfaces). It may be that questions of equilibrium are not to be settled by the ``moment-to-moment'' behaviour of three-dimensional integrals, but rather by global statements of the type represented by \cref{totdelNx}. This is something we are currently investigating and hope to be able to give more detail on in a future work.

\section{Perturbations with respect to equilibrium}\label{sec:PerturbativeExpansion}

With the equations of motion obtained from an action principle, we can consider perturbations away from equilibrium configurations (of the kind described above) in a way that is closely related---at least from the formal perspective---to standard hydrodynamical perturbation theory. The general approach to Lagrangian perturbation theory is perhaps best described by \citet{FriedmannSchutz1975}. Roughly speaking, the evolution equations for the perturbed fields can be obtained by perturbing the equations that follow from the action. It is also clear---at least in principle---how to construct a Lagrangian whose variation gives the perturbed equations (see \textsection 2 of \cite{FriedmannSchutz1975}). However, since  we are not focussing on a stability analysis of fluid oscillations we will not consider this additional aspect here. 

To set the stage for the perturbative expansion, we consider the family of worldlines (not necessarily geodesics) that each constituent of a multifluid system traces out in spacetime. Our definition of equilibrium includes the assumption that all species are comoving. Therefore, our fiducial set of worldlines representing equilibrium are those the system would have followed if it were comoving throughout its history.
\begin{figure}
    \centering
    \includegraphics[width=0.8\textwidth]{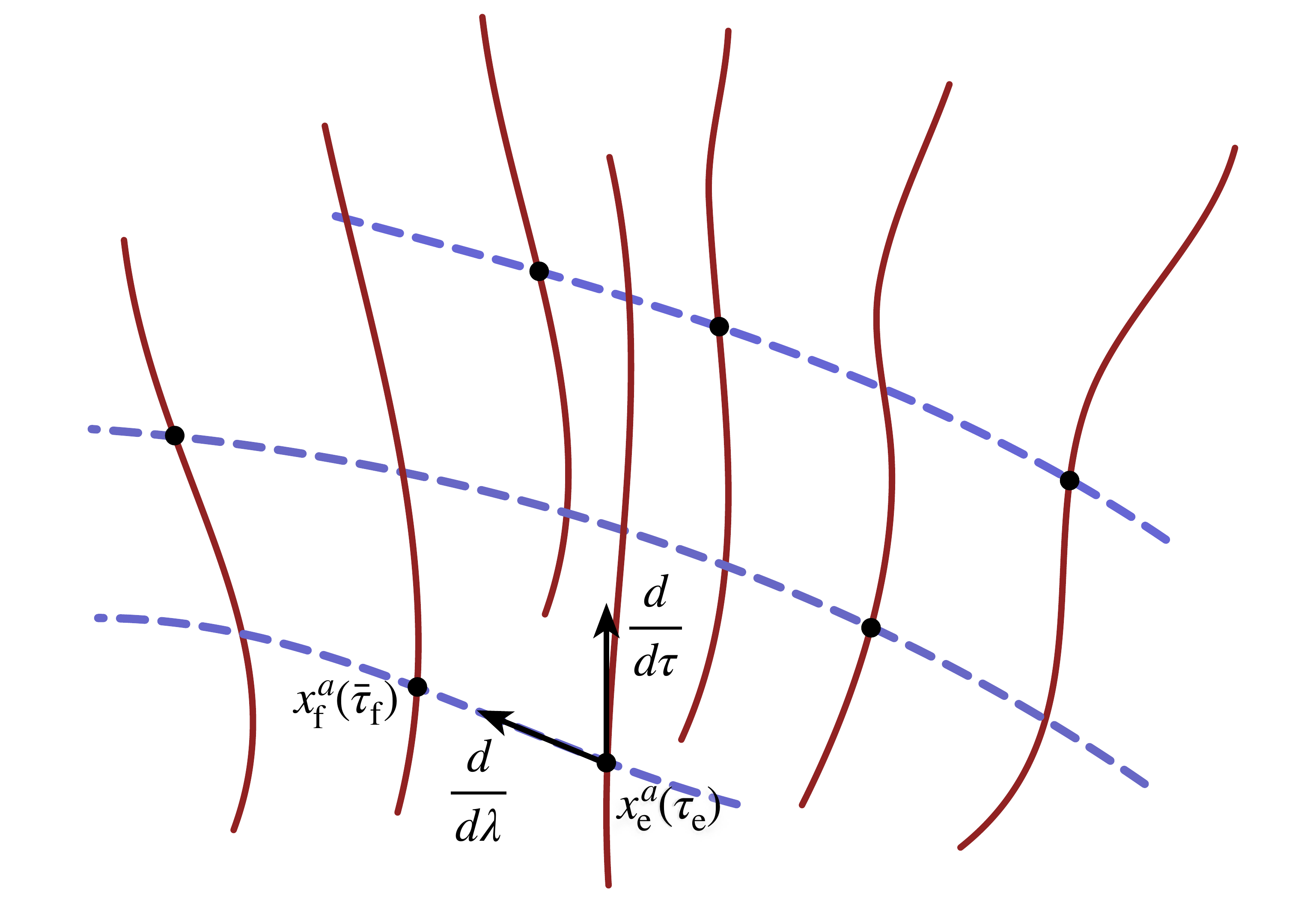}
    \caption{An illustration of worldlines associated with the fluid elements (solid vertical red lines, parameterized by $\tau,\,\bar\tau$) and “Lagrangian displacements” which connect fluid elements (dashed horizontal blue lines, parameterized by $\lambda$).}
    \label{fig:PerturbationWorldlines}
\end{figure}
This then allows us to view each of the ``final'' worldlines $x_\f^a(\bar \tau)$ as a curve in spacetime which is close to the equilibrium one $x_\e^a(\tau)$, with $\bar \tau$ and $\tau$ being the proper times of the respective curves. See \cref{fig:PerturbationWorldlines} for an illustration of the idea. The unit four-velocities associated with the two worldlines are
\begin{equation}
	u^a_\f = \frac{d x_\f^a}{d \bar \tau} \;, \qquad u^a_\e = \frac{d x_\e^a}{d \tau} \;.
\end{equation}
Obviously, $u^a_\e$ represents the comoving frame introduced earlier.

We assume another family of curves $x_{\e\f}^a(\lambda)$, where $\lambda$ is an affine parameter (say, the proper length), that connects the equilibrium worldline to the actual one. This means that for any point $x_\e^a(\tau_e)$ on the equilibrium worldline, there is a unique point $x_\f^a(\bar \tau_f)$ on the perturbed worldline, and a unique curve $x_{\e\f}^a(\lambda)$ between them having two points $x_{\e\f}^a\left(\lambda_\e\right)$ and $x_{\e\f}^a\left(\lambda_\f\right)$ such that
\begin{equation}
	x_\f^a\left(\bar \tau_\f\right) = x_{\e\f}^a\left(\lambda_\f\right) \;, \qquad x_\e^a \left(\tau_\e\right) = x_{\e\f}^a\left(\lambda_\e\right) \;.
\end{equation}
Taylor expanding the perturbed worldline about the equilibrium up to second order, we get
\begin{equation}
\begin{split}
	x_\f^a(\bar \tau_\f) &= x_\e^a(\tau_\e) + \frac{d x_{\e\f}^a}{d\lambda}\Big|_{\lambda_\e} (\lambda_\f - \lambda_\e) + \frac{1}{2} \frac{d^2 x_{\e\f}^a}{d^2\lambda}\Big|_{\lambda_\e} (\lambda_\f - \lambda_\e)^2 \\
 &= x_\e^a(\tau_\e) +\zeta^a \Delta\lambda + \frac{1}{2} \Big(\zeta^b\partial_b\zeta^a \Big)\Delta\lambda^2 \;,
\end{split}
\end{equation}
where we introduced the tangent vector 
\begin{equation}
	\frac{d}{d\lambda} = \frac{dx_{\e\f}^a}{d\lambda}\Big|_{\lambda_\e} \frac{\partial}{\partial x^a} = \zeta^a \partial_a \;.
\end{equation}

The first objects we want to perturb are the fluid element ``names''. That is, we attach a label $X^A$, where the index $A=1,2,3$, to each of the worldlines used to cover the region of spacetime occupied by the fluid. By definition of the Lagrangian variation \cite{FriedmannSchutz1975,Friedman1978a} we then have
\begin{equation}
	\Delta X^A = \Big(\phi_*  X^A(x_\f)\Big)(x_\e) - \bar X^A(x_\e) = X^A(x_\f) -\bar X^A(x_\e) = 0 \;, 
\end{equation}
where $\phi$ is the diffeomorphism that connects the perturbed and unperturbed worldlines, via the flow lines $x_{\e\f}^a$, and the last equality follows from the fact that the label does not change as we follow it. As a result we have, to first order 
\begin{equation}
	\delta X^A = -\mathcal{L}_{\xi_\x} X^A = -\xi_\x^a \Psi^A_{\e \,a} = - \xi_\x^A\;, 
\end{equation}
where we introduced the Lagrangian displacement vector $\xi^a_\x= x_\f^a - x_\e^a$. 

It is important to note that these displacement vectors are different from the ones introduced when obtaining the equations of motion from the action principle (see \cref{eq:ConvectiveVariation4Current}), even though the mathematics appears the same. In the present case the displacement vector connects two configurations that are ``close'' in the space of physical solutions---in field-theory parlance they are both ``on-shell''. We also note that, to compute the second order variation we cannot rely on the simple relation that exists between Lagrangian and Eulerian variation (at first order). We need to perform such calculations explicitly. 

At this point, it is worth pausing to consider what is behind the perturbation scheme we are building. Since we assume the existence of a well defined equilibrium timelike congruence $x_\e^a$ with four velocity $u_\e^a$, we may imagine  riding along with the equilibrium fluid element  observing the evolution of the system (towards equilibrium) from this perspective. This means that the x-species four-velocity $u_\x^a$ can be decomposed (in the usual way) as 
\begin{equation}
	u^a_\x = \gamma_\x \Big( u_\e^a + w_\x^a\Big) \;, \text{   where }\quad w_\x^a u_a^\e = 0 \;, \quad\gamma_\x = \Big(1 - w^a_\x w_a^\x\Big)^{-1/2} \;.
\end{equation}
Moreover, since we are working up to first order we have  
\begin{equation}
	\gamma_\x = 1 +\frac{1}{2} w_\x^2 \approx 1 + \mathcal{O}(2)  \quad \Longrightarrow \quad u_\x^a = u_\e^a + w_\x^a \;.
\end{equation}
We note that this linear expansion in the relative velocities, although in a different spirit, has also been discussed in the context of extensions to magneto-hydrodynamics \cite{ResistiveMultiMHDNils12,BeyondIdealMHD,BeyondMHDFibration}.

Also, it is interesting in itself (and necessary for perturbing the full set of fluid equations) to understand the relation between the spatial velocity $w_\x^a$ as measured by the equilibrium observer and the Lagrangian displacement $\xi^a_\x$. We consider the displacement to live in the local present of the equilibrium observer, i.e.,  to be such that $\xi_\x^a u^\e_a = \zeta_\x^au^\e_a =  0$.\footnote{This is essentially a gauge choice, see \cite{NilsGregReview} for discussion.} This implies that the vectors $\xi^a_\x$ and $\zeta^a_\x$ are spacelike non-null vector fields in spacetime. As a result, if we consider the proper time of the perturbed worldline, we have
\begin{equation}
	- d\bar \tau^2 = g_{ab}\,dx^a_\f\,dx^b_\f = g_{ab}\, dx_\e^a\,dx_\e^b + g_{ab} \Big( dx_\e^a\, \zeta^b \Delta \lambda + dx_\e^b\, \zeta^a \Delta \lambda\Big) = - d\tau^2 \;,
\end{equation}
where we used the fact that 
\begin{equation}
	x_\e^a = x_\e^a(\tau) \Longrightarrow dx_\e^a = u_\e^a d \tau \;.
\end{equation}
As a consequence, the proper time of the perturbed and equilibrium worldline is the same, so we have
\begin{equation}
	u_\x^a = \frac{dx_\f^a}{d \bar \tau} \approx \frac{dx_\e^a}{d\tau} + \frac{d}{d\tau} \xi^a_\x = u_\e^a + \dot\xi_\x^a
\end{equation}
where (again) the dot represents the covariant directional derivative in the direction of the equilibrium four-velocity.\footnote{To be more precise, one should distinguish between $\frac{d}{d\tau}=u^b_\e \partial_b$ and $\frac{D}{D\tau}=u^b_\e \nabla_b$. Since we are introducing a decomposition of a vector as a sum of two, $\dot\xi^a_\x$ must be a vector as well so that the dot represents a covariant directional derivative. } We observe that from the construction we have $w_\x^a = \dot\xi^a_\x$ and it is clear that when pushing the expansion to second order their relation will become more involved---both because the difference between the proper times ($\bar \tau$ versus $\tau$) appears at second order and because the Taylor expansion gets more complicated. 

We now aim to understand how to construct the expansion directly in matter space.  We start by noting that, since we are considering each displacement $\xi_\x^a$ to be orthogonal to $u^a_\e$ there is no loss of information in projecting the Lagrangian displacements onto the equilibrium matter space and dealing with $\xi^A_\x$. The general picture is thus as follows: in the general non-linear theory each matter space can be considered as an independent but interacting manifold, but this changes when we consider a perturbative expansion.  In fact, the fundamental assumption of perturbation theory is that the two configurations (perturbed and unperturbed) are related by some diffeomorphism. This implies that the perturbed and unperturbed matter spaces\footnote{Recall that the matter space is obtained by taking the quotient of the spacetime over the corresponding worldline, i.e. identifying the worldline as a single point.} are diffeomorphic, that is they are \textit{the same} abstract manifold. Therefore we can use the same chart on the two manifolds $X^A$ (label the worldlines in the same way) and the difference will be only in that $X_\x^A(x^a) \ne X_\e^A(x^a)$. The difference between the two will be exactly what we found above, namely $-\xi_\x^A$. We also note that, by our definition of the unperturbed state, all the perturbed matter spaces are diffeomorphic to the same unperturbed one, and thus to each other. 

Given this, we can work out how a general matter space tensor transforms under diffeomorphisms \citep{GRCarroll}. For instance, if we consider the projected metric $g_\x^{AB}$ we have\footnote{For the Lie derivative we use the formula with partial derivatives in order to avoid the possible confusion arising from the choice of the connection used on the matter space.}
\begin{equation}
	\delta g_\x^{AB} = -\mathcal{L}_{-\xi_\x}g_\x^{AB} = \mathcal{L}_{\xi_\x} g_\x^{AB} = \xi_\x^C \partial_C g_\e^{AB} - g_\e^{CB}\partial_C\xi^A_\x - g_\e^{AC}\partial_C\xi_\x^B \,
\end{equation}
where the partial derivatives are taken with respect to the equilibrium matter space coordinates. We now observe that, considering $\xi^A_\x$ as a scalar field in spacetime we can write
\begin{equation}
	-g_\e^{CB} \partial_C\xi^A = - g^{ab} \Psi^C_{\e\,a} \Psi^A_{\e\,b}\partial_C\xi_\x^A = - \Psi^A_{\e\,a} \nabla^a \xi^A_{\x} \;.
\end{equation}
We also note that, since\footnote{If this is not immediately convincing one can prove it by taking the explicit definition of a derivative on the coordinate functions $X^A(\bar X) = \delta^{A}_{\,C}\bar X^C = \bar X^A$ and using the linearity of the derivative.} $\partial_C \Psi^A_{\e\,a} = \partial_a \delta^A_C = 0,$ we have
\begin{equation}
	\partial_C g_\e^{AB} = 2\, g^{ab}\Bigg(\frac{\partial}{\partial X_\e^C} \Psi^A_{\e\,a}\Bigg) \Psi^B_{\e\,b} = 0 \;.
\end{equation}
As a result, the projected metrics transform as
\begin{equation}\label{eq:VariationGx}
	\delta g_\x^{AB} = - \Psi^B_{\e\,a} \nabla^a\xi^A_{\x} - \Psi^A_{\e\,a}  \nabla^a \xi^B_{\x} \;.
\end{equation}
This also tells us that building the variation of the metric tensor in this way, we are only comparing  the difference in the position of the particles, keeping fixed the spacetime metric.

We can now use the definition in \cref{eq:ProjectedDerivative} to decompose the displacement gradients as
\begin{equation}
	\nabla_a\xi_{\x}^A = -w_\x^A u^\e_a + D_a\xi_\x^A 
\end{equation}
and rewrite 
\begin{equation}\label{eq:DeltaMetricGradientsDisplacements}
\begin{split}
	\delta g^{AB}_\x &= \Psi^B_{\e\,a} (w_\x^A u^a_\e - D^a\xi_\x^A ) + \Psi^A_{\e\,a} (w_\x^B u^a_\e - D^a \xi_\x^B ) \\
    &= - D^B\xi_\x^A - D^A \xi_\x^B \;,
\end{split}
\end{equation}
where we introduced the short-hand notation $D^A = \Psi^A_{\e\,b}\,g^{ab}D_a$. It is  worth noting that \cref{eq:DeltaMetricGradientsDisplacements} is not a strain-rate tensor of the type usually introduced in fluid dynamics, because it involves gradients in the displacements instead of velocities. The usual strain rate tensor is in fact\footnote{To see this one has to use  $\mathcal{L}_{u_\x}\Psi^A_{\x\,a}=0$.}
\begin{equation}\label{eq:DotGAB}
\begin{split}
	\dot g_\x^{AB} &= -2\, \Psi^A_{\x\, (a}\Psi^B_{\x \,b)} \big[  -u_\x^b\dot u^a_\x + \varpi^{ab}_\x+ \sigma^{ab}_\x + \frac{1}{3}\theta_\x \perp^{ab}_\x\big] = \\ 
    & = - 2 \, \Psi^A_{\e \,(a}\Psi^B_{\e\, b)} \big(\sigma_\x^{ab} + \frac{1}{3}\theta_\x \perp^{ab}_\e\big)  + \mathcal{O}(2) = -2 \big(\sigma_\x^{AB} + \frac{1}{3}\theta_\x g^{AB}_\e\big)  \; ,
\end{split}
\end{equation}
We will comment on the implications of this difference later. 

Even if  it is not entirely obvious what kind of object the mixed projected metric $g^{AB}_{\x\y}$ is in the general non-linear case, in the context of a perturbative expansion there is no real difference between the various matter spaces (they are all diffeomorphic to the equilibrium one).  This means that we can use the same fundamental formula also for $g_{\x\y}$ to get
\begin{equation}\label{eq:VariationGxy}
\begin{split}
	\delta g_{\x\y}^{AB} &= g_{\x\y}^{AB} - g_\e^{AB} = g^{ab} \Big( \delta \Psi^A_{\x\,a} \Psi^B_{\e\,b} + \delta \Psi^B_{\y\,b} \Psi^A_{\e\,a}\Big) = \\
    & = - \Psi^B_{\e\,a} \nabla^a \xi^A_{\x} - \Psi^A_{\e\,a}\nabla^a \xi^B_{\y}  \;.
\end{split}
\end{equation}
It is interesting to note that since $\delta g^{ab} = 0$ we
have
\begin{equation}
	[\delta, \nabla_a] = [\delta , \partial_a ] = 0 \;.
\end{equation}
That is, the variation commutes with both partial and covariant derivatives. This will become relevant when we need to work out the variation of the resistive terms that stem from a dependence of the $\N_\x$ on $g_{\x\y}^{AB}$ and $g_\y^{AB}$. 

There has been a number of recent efforts on building first-order dissipative hydrodynamic models starting from a field-theory perspective. It makes sense to point out the differences between the present expansion and the field-theory-based ones. From a field theory perspective hydrodynamics is the low-energy limit of a more fundamental theory. Starting from this point of view, different authors have proposed (see, for example, \cite{Kovtun19,Bemfica19}) to introduce dissipation in the models through a gradient expansion. Practically, this boils down to postulating the most general constitutive equations---that is the relations between thermodynamical forces and fluxes---in terms of the standard hydrodynamical variables (like $T,\,\mu\dots$) and their derivatives. In this context, the models are said to be of first order if the constitutive equations involve all permissible terms with just one derivative. When the system is close to equilibrium one can expect the gradients in temperature, chemical potential etc$\dots$ to be small, so that terms with two or more derivatives are dominated by first-order ones. The final aim is (again) to obtain a set of equations valid close to equilibrium. 

In contrast, in the present work, the variables that define the physical state of the system take values close to the equilibrium ones, and by ``first order'' we mean the deviations are expanded up to $\mathcal{O}(\xi_\x)$. It is therefore clear that the present approach differs from the field-theory-based (gradient) expansions. The ultimate reason is that the action-based model provides the exact equations, which we then approximate, while in the field-theory approach one is trying to build the full equations as successive expansions. 

\section{Energy Density is Stationary at Equilibrium}\label{sec:EnergyMinimized}
In order to describe out-of-equilibrium systems with the Extended Irreversible Thermodynamic (EIT) paradigm \citep{EIT}, one postulates the existence of a generalized entropy---a function of a larger set of Degrees of Freedom than the corresponding equilibrium ones---which is maximized at equilibrium. The starting point for the formalism used here is a generalized energy where the only degrees of freedom are the fluxes. The action-based model provides the total stress-energy-momentum tensor $T_{ab}$ of the system, so that we can easily extract the total energy density $\veps$ for some observer having four-velocity $u^a$ via the projection $\veps = u^a u^b T_{a b}$. We will now show that requiring the local energy density to be at a minimum in equilibrium means the viscous stress tensors have to be zero. 

When specific modeling is carried out, such as a numerical evolution, we would need to provide an equation of state (EoS) and specify values for the microphysical input parameters. From the phenomenological point of view, this corresponds to assuming the existence of a function---in our case, energy density---defined on some ``thermodynamical manifold'' whose coordinates are the relevant degrees of freedom. Practically speaking, the formalism developed here suggests we may identify the thermodynamical manifold with the matter space used in the variational model. As the general discussion gets quite complex, we focus on the specific example of a two-component system, with the components representing matter and entropy (see \cite{Lopez2011,NilsHeat2011}). 

Let us first consider the non-dissipative limit.  Thermodynamics of a single fluid is described by some equilibrium energy $\varepsilon_\e(n,s)$ such that 
\begin{equation}
	d\varepsilon_\e = T ds + \mu dn = \sum_{\x=\n,\s}\mu^\x dn_\x \;.
\end{equation}
On the other hand, the conservative variational model is built using a master function $\Lambda(n_\n^2, n_\s^2, n_{\n\s}^2)$. Because of our assumption that all species are comoving in equilibrium there is no heat flux relative to the matter and therefore $n_{\n\s}^2 = - g_{ab} n_\n^a n_\s^b = + n_\n n_\s$, and the master function only depends on two variables, $\Lambda = \Lambda(n_\n,n_\s)$. It is indeed easy to see that the equilibrium energy density, as measured by the equilibrium observer, is
\begin{equation}
	\hat\varepsilon_\e = T_{ab}\,u_\e^a u_\e^b = \big[\Psi g_{ab} + (\Psi - \Lambda)u^\e_a u^\e_b\big] u_\e^a u_\e^b = - \Lambda
\end{equation}

Since we have already identified the matter space normalizations of the three-forms with the rest frame densities $\N_\x = n_\x$, we can think of the thermodynamic energy as a function defined on the matter space, and write
\begin{equation}
	\hat\varepsilon_\e = \hat\varepsilon_\e(\N_\n,\N_\s) = - \Lambda_\e(\N_\n,\N_\s)
\end{equation}
The equilibrium case suggests that we could try to extend this identification to the non-equilibrium setting, and ``build'' the thermodynamics on the matter space. This raises the (difficult) question of what the global matter space is in the full non-linear case. We  will not address that issue here. Instead, we focus on the near-equilibrium case, where we only have to deal with the equilibrium matter space. 

Because of the way we have built the expansion, it is natural to project tensor quantities---fluxes, stress-energy-momentum tensor, etcetera---into the frame of the equilibrium observer, as defined by the equilibrium worldlines congruence $u^a_\e$. Quantities  measured in this frame will be indicated by a ``hat'' in the following. Objects without a hat are measured in fluid rest frames, which are defined by the $u^a_\x$. The equilibrium value of a quantity in the equilibrium frame will be indicated with a ``bar''. For instance, the particle density measured in the equilibrium frame is $\hat n_\x = - u^\e_a n^a_\x$; in the $\x$-fluid rest frame it is $n_\x = - u^\x_a n^a_\x$; and the equilibrium value in the equilibrium frame is $\bar n_\x = \hat n_\x\big|_\e$.

The ``out-of-equilibrium'' energy density $\hat\varepsilon_{\o.\e.}$ of the system as determined in the equilibrium rest frame is given by 
\begin{equation}
	\hat\varepsilon_{\o.\e.} = \big( T^{ab}_{\n.\d.} + \sum_\x D^{ab}_\x \big) u^\e_a u^\e_b = \varepsilon_{\o.\e.}^{\n.\d.} + D^{ab} u^\e_a u^\e_b \;,
\end{equation}
where we have separated the contribution from the viscous stress tensor $D_{ab}$ from those having the ``non-dissipative'' form
\begin{equation}
	T^{ab}_{\n.\d.} = \Big(\Lambda - \sum_\x n_\x^c \mu^\x_c\Big) g^{ab} + \sum_\x n_\x^a\mu_\x^b = \Psi\,g^{ab}+ \sum_\x n_\x^a\mu_\x^b \; .
\end{equation}
The expression for $\hat\varepsilon_{\o.\e.}$ can be made more explicit by means of \cref{muxcomov}, which leads to $\Psi = \Lambda + \sum_\x n_\x \mu_\x$ and
\begin{equation}
	\hat\varepsilon_{\o.\e.}^{\n.\d.} = u^\e_a u^\e_b T^{ab}_{\n.\d.} = - \Lambda - \sum_{\x= n,s} \big(n_\x \mu_\x - \hat n_\x \hat\mu_\x\big) \;.
\end{equation}

Because the flux is a vector, the two densities $\hat n_\x$ and $n_\x$ are easily shown to be related by
\begin{equation}\label{eq:restvseqdensity}
	\hat n_\x = - n_\x^a u_a =  -n_\x u_\x^a u_a = (1 - w_\x^aw^\x_a)^{-1/2} n_\x = \Big( 1 + \frac{1}{2}w_\x^2\Big)n_\x + \mathcal O(3) \;.
\end{equation}
Meanwhile, the corresponding momentum relation is a bit more involved because of entrainment: 
\begin{equation}
\begin{split}
		\mu_\x &= -\mu^\x_b u_\x^b = -\gamma_\x (u^b + w_\x^b) \big(\mathcal B_\x n_\x u^\x_b + \sum_{\y \neq \x} \mathcal A_{\x\y} n_\y u_\y^b\big) \\
        &= \gamma_\x \Big( \hat\mu_\x - \mathcal B_\x n_\x \gamma_\x w_\x^2 -  \sum_{\y \neq \x} \mathcal A_{\x\y} n_\y \gamma_\y w_\x^aw^\y_a\Big) \;.
\end{split}
\end{equation}
We can rearrange this as 
\begin{equation}\label{eq:restvseqmomentum}
	\hat\mu_\x = \mu_\x - \frac{1}{2}\bar\mu_\x w_\x^2 + \bar {\mathcal B}_\x \bar n_\x w_\x^2 + \sum_{\y\neq \x} \mathcal{\bar A }_{\x\y}\bar n_\y w_\x^aw^\y_a 
\end{equation}
and, wrapping up, we get 
\begin{equation}\label{eq:nondissenergylambda}
\begin{split}
	\hat\varepsilon_{\o.\e.}^{n.d.} &= -\Lambda  + \bar {\mathcal B}_\n \bar n_\n^2 w_\n^2 +\bar {\mathcal B}_\s \bar n_\s^2 w_\s^2 + 2\bar \A_{\n\s}\bar n_\s\bar n_\n w_\n^aw^\s_a \\
    &=  -\Lambda +\bar\mu_\n\bar n_\n w_\n^2 +\bar\mu_\s\bar n_\s w_\s^2 - \A_{\n\s}\bar n_\n \bar n_\s w_{\n\s}^2\;,
\end{split}
\end{equation}
where 
\begin{equation}
 w_{\x\y}^2 = g_{a b} \big(w_\x^a - w_\y^a\big) \big(w_\x^b - w_\y^b\big)   \;. 
\end{equation}
It is now clear that, in order to proceed, we need an expansion for the master function, $\Lambda$. 

Note that the dissipative action model assumes $\Lambda$ depends on $(X_\n^A, X_\s^A, g_\n^{AB} , g_\s ^{AB}, g_{\n\s}^{AB})$ through the scalar product of the fluxes $n_\n^2, n_\s^2, n_{\n\s}^2$. Therefore, we can expand $\Lambda$ up to second order in the standard way (see \cite {NilsGregReview}). We thus have 
\begin{equation}\label{eq:Master2Ord}
\begin{split}
	\Lambda  = \Lambda_{\e} &-\frac{1}{2} \sum_{\x=n,s}  \B_\x\delta n_\x^2 - \A_{\n\s} \delta n_{\n\s}^2 - \frac{1}{4} \sum_{\x=n,s} \frac{\partial \B_\x}{\partial n_\x^2} (\delta n_\x^2)^2 -\frac{1}{2}\frac{\partial\A_{\n\s}}{\partial n_{\n\s}^2}  (\delta n_{\n\s}^2)^2 \\ 
    &-\frac{1}{2}\frac{\partial\B_\n}{\partial n_{\s}^2} (\delta n_\n^2)(\delta n_\s^2 )- \frac{\partial\A_{\n\s}}{\partial n_\n^2}(\delta n_\n^2)(\delta n_{\n\s}^2) - \frac{\partial\A_{\n\s}}{\partial n_\s^2} (\delta n_\s^2)(\delta n_{\n\s})^2 \;.
\end{split}
\end{equation}
To make contact with the previous expansion on the matter space we need  explicit expressions for $\delta n_\x^2$ and all  other similar terms that appear in this expression. 

For the four-current we have
\begin{equation}\label{eq:deltanexa}
\begin{split}
	\delta n_\x^a &= n_\x^a - \bar n_\x^a = (\bar n_\x + \delta n_\x )\Big[\big(1+\frac{1}{2}w_\x^2\big)u^a + w_\x^a\Big] - \bar n_\x u^a \\
    &= \frac{1}{2} \bar n_\x w_\x^2u^a + \bar n_\x w_\x^a +\delta n_\x u^a + \delta n_\x w_\x^a \;,
\end{split}
\end{equation}
and we see that it---quite intuitively---changes both as the density and the four-velocity change. By means of \cref{eq:deltanexa} we get
\begin{equation}\label{eq:deltanex2}
	\delta n_\x^2 = - \big(2\bar n_\x^a \delta n^\x_a+ \delta n_\x ^a \delta n^\x_a\big)  = 2 \bar n_\x \delta n_\x + (\delta n_\x)^2 \;.
\end{equation}
Similarly, we have
\begin{equation}\label{eq:deltanexway2}
\begin{split}
	\delta n_{\x\y}^2 &= - \Big( \bar n_\x^a \delta n_a^\y + \bar n_\y^a \delta n^\x_a + \delta n_\x^a \delta n^\y_a \Big)  \\
    &= \bar n_\x \delta n_\y+ \bar n_\y \delta n_\x + \delta n_\x \delta n_\y + \frac{1}{2} \bar n_\x \bar n_\y w_{\x\y}^2 \;.
\end{split}
\end{equation}

In order to complete the second order expansion of $\Lambda$ we also need the products (for every possible combination) of \cref{eq:deltanex2} and \cref{eq:deltanexway2}. These are found to be
\begin{subequations}
\begin{align}
	\big(\delta n_\x^2\big)^2  &= 4 \bar n_\x^2 (\delta n_\x)^2 \;,\\
      (\delta n_{\x\y}^2)^2 &= \bar n_\x^2 (\delta n_\y)^2 + \bar n_\y^2 (\delta n_\x)^2+ 2\bar n_\x \bar n_\y \delta n_\x \delta n_\y \;,\\
     (\delta n_\x^2) (\delta n_\y^2) &= 4 \bar n_\x \bar n_\y \delta n_\x \delta n_\y \;,\\
     (\delta n_{\x\y}^2)( \delta n_\x^2) &= 2\bar n_\x (\delta n_\x) \big(\bar n_\y \delta n_\x + \bar n_\x \delta n_\y\big)\;.
\end{align}
\end{subequations}
Plugging these expressions into \cref{eq:Master2Ord} we find (up to second order) 
\begin{equation}\label{eq:2OrdOffEqEnergy}
\begin{split}
	\hat\varepsilon^{n.d.}_{\o.\e.} &= \varepsilon_\e(\bar n_\n, \bar n_\s) + \bar \mu_\n \delta n_\n + \bar \mu_\s \delta n_\s  + \frac{1}{2} \big(\bar\B_\n \bar c_\n^2 -\bar\A^{\n\n}_{uu} \big)( \delta n_\n)^2 \\
    &+ \frac{1}{2} \big(\bar\B_\s \bar c_\s^2 -\bar\A^{\s\s}_{uu} \big)( \delta n_\s)^2 - \big(\bar\chi^{\s\n}_{uu} + \bar\A^{\n\s}_{uu} \big) (\delta n_\n)(\delta n_\s) + \bar\mu_\n\bar n_\n w_\n^2 \\
    &+ \bar\mu_\s\bar n_\s w_\s^2   - \frac{1}{2}\bar\A_{\n\s}\bar n_\n \bar n_\s w_{\n\s}^2  \;,
\end{split}
\end{equation}
where we have made use of \cref{eq:nondissenergylambda} and defined
\begin{subequations}\label{eq:SecondOrderMasterUsefulDef}
\begin{align}
	\bar c_\x^2 &= 1 + 2\frac{\bar n_\x^2}{\bar\B_\x}\frac{\partial\bar\B_\x}{\partial n_\x^2} \\
  \bar  \A^{\x\x} _{ab} &= -\Big(\bar n_\y^2 \frac{\partial\bar\A_{\x\y}}{\partial n_{\x\y}^2} + 4 \bar n_\x \bar n_\y \frac{\partial\bar\A_{\x\y}}{\partial n_\x^2}\Big) u^\e_a u^\e_b \doteq \bar  \A^{\x\x} _{uu} u^\e_a u^\e_b\\
  \begin{split}
  \bar  \A^{\n\s}_{ab} &= \bar\A^{ns}\perp_{ab} -\Big(\bar\A_{\n\s} + 2\bar n_\n^2 \frac{\partial\bar\A^{\n\s}}{\partial n_\n^2} + 2\bar n_\s^2 \frac{\partial\bar\A^{\n\s}}{\partial n_\s^2} + \bar n_\n\bar n_\s \frac{\partial\bar\A^{\n\s}}{\partial n_{\n\s}^2}\Big) u^\e_a u^\e_b \\
  &\doteq\bar\A^{\n\s}\perp_{ab} + \bar  \A^{\n\s}_{uu} u^\e_a u^\e_b
  \end{split} \\
    \bar\chi^{\n\s}_{uu} &= -2\bar n_\n\bar n_\s \frac{\partial\bar\B^\n}{\partial n_\s^2} = -2\bar n_\n\bar n_\s \frac{\partial\bar\B^\s}{\partial n_\n^2} 
\end{align}
\end{subequations}
See \citet{NilsGregReview} for more discussion of these terms and \citet{samuelsson10:_rel2st} for their roles in the context of two-stream instability.

Noting that the quantity $\delta n_\x$ is the variation of the rest frame density, we can relate it to a variation of $\N_\x$ and ``close the loop''. Since the $\N_\x$ are functions on matter space of the variables $(X_\n,\;X_\s,\;g^{AB}_\n,\;g^{AB}_\s,\;g^{AB}_{\n\s})$ the expression for the energy is actually a second order expansion in terms of those variables. We note also that, because of the two-layer structure, the $\delta n_\x$ above contain second-order terms.

A priori, the expression in \cref{eq:2OrdOffEqEnergy} does not provide the total out-of-equilibrium energy because we also need to account for the dissipative terms. However, we will now show that these actually do not contribute. To do this, we assume an expansion for all the viscous stress tensors of form 
\begin{equation}
	S_{AB} = S^{\e}_{AB} + S_{AB}^1 + S^2_{AB} + \mathcal{O}(3)
\end{equation}
without providing (for now) the explicit expressions. Recalling \cref{eq:MatterSpacelinearMap} and $\Psi^A_{\e\,a} u^a_\e = 0$, we can write
\begin{equation}
	S_{ab} u^a_\e u^b_\e = S_{AB} (X_{\e}^A + \delta X^A)_{,a}(X_{\e}^B + \delta X^B)_{,b} u^a_\e u^b_\e = S_{AB}^{\e}\,\delta X^A_{\;,a}\,\delta X^B_{\;,b} u^a_\e u^b_\e \;,
\end{equation}
where the expansion is up to second order. It is clear that this argument is valid for each viscous stress tensor, and for $D_{ab} u^a_\e u^b_\e$ as well, so that the dissipative contributions to the off-equilibrium energy are, at least, of second order. Assuming that the energy is stationary, that is
\begin{equation}
	\hat\varepsilon^{n.d.}_{\o.\e.} -  \varepsilon_\e(\bar n_\n, \bar n_\s) = 0 + \mathcal{O}(2) \;,
\end{equation}
we then have
\begin{equation}\label{eq:MinimumeEnergyCondition}
	\bar\mu_\n \delta n_\n + \bar\mu_\s\delta n_\s = \mathcal{O}(2) \;,
\end{equation}
which has a clear thermodynamical interpretation and is consistent with the EIT picture, see \citep{EIT}, since, up to first order, the generalized energy is a function of the $n_\x$ only. 

We want to translate the above result into conditions for the matter space functions $\N_\x$. We start by observing that in the conservative case, the three-form $n_{ABC}^\x$ is a function of the $X_\x^A$ coordinates only. Therefore, $\bar\N_\x$ is just a function of $X_\x^A$, while, because  $\N_\x = \bar\N_\x\sqrt{g^\x}$, the latter depends also on the projected metric 
\begin{equation}
	\frac{\partial \N_\x}{\partial g_\x^{AB}} = \frac{1}{2} \sqrt{g^\x}\bar\N_\x g^\x_{AB} = \frac{1}{2}\N_\x g^\x_{AB} \;.
\end{equation}
When considering the expansion of $n_\x$ (and hence $\N_\x$) we assume that we can write 
\begin{equation}
	\N_\x = \N_\x^\e + \N_\x^\d \;,
\end{equation}
where $\N_\x^\e$ is the same as in the non-dissipative limit while the dissipative contribution $\N_\x^\d$ is a function also of the additional variables that encode the dissipation. Given the separation of $\N_\x$ into two pieces it is natural to assume that $\N_\x^\d$, but not its derivatives, vanishes at equilibrium. 

Since the equilibrium evolves in a conservative fashion, we can write
\begin{equation}
\begin{split}
		\delta n_\x &\equiv \N_\x - \N_\x^\e = \N_\x^\d = \N_\x^\d -\N_\x^\d\Big|_{\e} \\
    &= \frac{\partial \N_\x^\d}{\partial X_\x^A} \delta X_\x^A +  \frac{\partial \N_\x^\d}{\partial X_\y^A} \delta X_\y^A +  \frac{\partial\N_\x^\d}{\partial g_\x^{AB}} \delta g_\x^{AB} + \frac{\partial\N_\x^\d}{\partial g_\y^{AB}} \delta g_\y^{AB} + \frac{\partial \N_\x^\d}{\partial g_{\x\y}^{AB}} \delta g_{\x\y}^{AB} +\mathcal{O}(2) \;,
\end{split}
\end{equation}
where here, and in similar expansions below, each quantity is to be evaluated at equilibrium. With this assumption it is easy to read off from \cref{eq:MinimumeEnergyCondition} the first order relation
\begin{equation}\label{eq:1stOrderEquilCondition}
    	\M_\n \delta\N_\n^\d+ \M_\s \delta\N_\s^\d =0 \ .
\end{equation}
This leads to
\begin{equation}
  	\M_\n \frac{\partial\N_\n^\d}{\partial X_\n^A} + \M_\s \frac{\partial\N_\s^\d}{\partial X_\n^A} =0 \ , 
\end{equation}
and analogous results for variations with respect to $X_\s^A$, $g_\n^{AB}$, $g_\s^{AB}$ and $g_{\n\s}^{AB}$ follow immediately.
In particular, this shows that the total viscous stress tensor, acting on each component $D^\x_{ab}$, vanishes when the energy is minimized. 

To see this explicitly we note that (see \cref{eq:StdViscousTensorSlim,eq:AdditionalViscousTensorsSlim})
\begin{equation}\label{eq:mixedviscousdie}
	\mathcal{S}^{\x\y,\,\e} _{AB} \equiv 2\M_\x\frac{\partial\N_\x}{\partial g_{\x\y}^{AB}}= - 2\M_\y\frac{\partial\N_\y}{\partial g_{\x\y}^{AB}}= -\mathcal{S}^{\y\x,\,\e}_{BA}
\end{equation}
where we made use of the symmetry property of the mixed metric, namely $g_{\x\y}^{AB} = g_{\y\x}^{BA}$. 
Similarly,
\begin{equation}\label{eq:viscousdie}
\begin{split}
	S^{\x,\,\e}_{AB} & \equiv 2 \M_\x\Big(\frac{\partial \N_\x}{\partial g_\x^{AB}} -\frac{1}{2} \N_\x g^\x_{AB}\Big)  \\
    &=  2 \M_\x\Big[\frac{\partial \N_\x^\d}{\partial g_\x^{AB}} -\frac{1}{2} \big( \N_\x -\N_\x^\e \big)g^\x_{AB}\Big]  \\
    &=2 \M_\x \frac{\partial \N_\x^\d}{\partial g_\x^{AB}}= -  2 \M_\y\frac{\partial \N_\y^\d}{\partial g_\x^{AB}}= - s^{\y\x,\,\e}_{AB}
\end{split}
\end{equation}
It is now clear that, by means of \cref{eq:mixedviscousdie} and \cref{eq:viscousdie}, the x-species viscous stress tensor vanishes:
\begin{equation}
	D^{\x,\,\e}_{AB} = S^{\x,\,\e}_{AB} + s^{\y\x,\,\e}_{AB} +\frac{1}{2}(\mathcal{S}^{\x\y,\,\e}_{AB} + \mathcal{S}^{\y\x,\,\e}_{BA}) =0\;.
\end{equation}

We have considered the fully general case with all the additional dependences in $\N_\x$ and all the viscous tensors $S^\x_{ab}$, $\mathcal{S}^{\x\y}_{ab}$ and $s^{\x\y}_{ab}$. The same result---that is, each $D^{\x,\,\e}_{ab}$ vanishes---holds even in a less rich situation when the model is built from fewer viscous tensors. In that case we have to go back to \cref{eq:1stOrderEquilCondition} and modify the argument accordingly. It is important to stress that we have shown that the full stress-energy-momentum tensor at equilibrium is made out of just the non-dissipative part, and that the dissipative parts of the total stress-energy-momentum tensor do not contribute to the total energy density at second order. 

However, we note that the energy minimum conditions in \cref{eq:1stOrderEquilCondition} do not set the purely resistive terms to zero (\cref{eq:PurelyReactiveSlim}). In fact, it only leads to 
\begin{equation}
\begin{split}
	&\M_\n\frac{\partial \N^\d_\n}{\partial X_\n^A} = -\text R^{\s\n,\,\e}_A \;,\\
    &\M_\s\frac{\partial \N^\d_\s}{\partial X_\s^A} = -\text R^{\n\s,\,\e}_A \;.
\end{split}
\end{equation}
The reason for this is pretty clear as these terms do not enter the expression for the energy density. We nonetheless might want to consider the case where the equilibrium equations are exactly as the conservative ones. The motivation for this can be found in the derivation of the purely resistive terms. If the different species are comoving at the action level, there is no distinction between the different $X_\x^A$ and no resistive term of this form would appear. We can enforce consistency with this observation in two ways: either we assume that we use the complete dependence on $X_\x^A$ in the conservative part, in which case
\begin{equation}
	\frac{\partial  \N^\d_\x}{\partial X_\x^A}\Big|_{\e} =0  \Longrightarrow \text R^{\x\y,\,\e}_A = 0 
\end{equation}
or, we just set the terms $\bar{\text R}^\x_a$ to zero, so that
\begin{equation}
	\bar\M_\n\frac{\partial \N^\d_\n}{\partial X_\n^A}\Big|_{\e} = \bar\M_\s\frac{\partial \N^\d_\s}{\partial X_\s^A}\Big|_{\e} \;.
\end{equation}
The latter, less restrictive assumption reminds us of  the dynamical nature of chemical equilibrium in nature. Reactions happen also at equilibrium, although they do so in such a way that there is no net particle production. Such equilibrium reactions are key to explaining neutron star cooling.

Finally, it is  quite easy to see that if we choose a different observer, such as the ones associated with the Eckart or Landau frame, the differences in the energy density will be of second order. Crucially, the equilibrium conditions in \cref{eq:1stOrderEquilCondition} do not depend on the choice of frame.

\section{The last piece of the puzzle}\label{sec:LastPiece}

In order to work out the perturbative expressions we need to expand the various dissipative terms. It should now be clear that for the viscous stress tensors we can write\footnote{All the derivatives are intended to be evaluated at equilibrium.}
\begin{subequations}\label{eq:ExpansionViscousTensorsMatter}
\begin{align}
	\delta s^{\x\y}_{AB} &= 2\frac{\partial\N_\x^\d}{\partial g_\y^{AB}}\delta \M_\x 	+ 2\bar\M_\x \delta \Bigg(\frac{\partial\N_\x^\d}{\partial g_\y^{AB}}\Bigg) \;, \\
    \delta \mathcal{S}^{\x\y}_{AB} &= 2\frac{\partial\N_\x^\d}{\partial g_{\x\y}^{AB}}\delta \M_\x  + 2\bar\M_\x\delta \Bigg(\frac{\partial\N_\x^\d}{\partial g_{\x\y}^{AB}}\Bigg) \;,\\
    \delta S^{\x}_{AB} &= 2\frac{\partial\N_\x^\d}{\partial g_\x^{AB}}\delta \M_\x 	+ 2\bar\M_\x\delta \Bigg(\frac{\partial\N_\x^\d}{\partial g_\x^{AB}}\Bigg) -\bar\M_\x (\delta \N_\x^\d) g_{AB}^\e \;,
    \end{align}
\end{subequations}
where
\begin{subequations}
\begin{align}
    & s^{\x\y}_{AB} = 2\M_\x \frac{\partial \N_\x^\d}{\partial g_\y^{AB}} \;,\\
    &\mathcal{S}^{\x\y}_{AB} = 2 \M_\x \frac{\partial\N^\d_\x}{\partial g_{\x\y}^{AB}} \;,\\
    &S^\x_{AB} = 2\Bigg( \M_\x \frac{\partial \N_\x^\d}{\partial g_\x^{AB}} - \frac{1}{2}\M^\x\N_\x^\d\,g^\x_{AB} \Bigg) \;.
\end{align}
\end{subequations}
Similarly, for the ``purely resistive'' terms we have 
\begin{equation}\label{eq:ExpansionPurelyReactiveMatter}
	\delta \text R^{\x\y}_A =  \frac{\partial \N_\x^\d}{\partial X_\y^{A}}\delta \M_\x	+ \bar\M_\x\delta \Bigg(\frac{\partial \N_\x^\d}{\partial X_\y^{A}}\Bigg) \;.
\end{equation}
Since $\N_\x^d$ is a function of $(X_\x,\,X_\y,\,g^{AB}_\x,\,g^{AB}_\y,\,g^{AB}_{\x\y})$, its derivatives are as well, so that we have
\begin{equation}\label{eq:PertDerivBarN}
\begin{split}
	\delta \Bigg(\frac{\partial\N_\x^\d}{\partial X_\y^A}\Bigg) = &\frac{\partial ^2 \N_\x^\d}{ \partial X_\x^B \partial X_\y^A} \delta X^B_\x +  \frac{\partial ^2 \N_\x^\d}{ \partial X_\y^B \partial X_\y^A} \delta X^B_\y +\frac{\partial ^2 \N_\x^\d}{ \partial g_\x^{BC} \partial X_\y^A}\delta g_\x^{BC} \\
    &+\frac{\partial ^2 \N_\x^\d}{ \partial g_\y^{BC} \partial X_\y^A} \delta g_\y^{BC} +\frac{\partial ^2 \N_\x^\d}{ \partial g_{\x\y}^{BC} \partial X_\y^A} \delta g_{\x\y}^{BC} \;.
\end{split}
\end{equation}
Similar results hold for the other variations that were not explicitly written in \cref{eq:ExpansionViscousTensorsMatter} and \cref{eq:ExpansionPurelyReactiveMatter}. 

Concerning the purely resistive term we note that $u_\x^a\text R^{\y\x}_a=0$ automatically. Because we are doing an  expansion with undetermined coefficients, we need to impose this by hand at every order; specifically, at the linear level. This then leads to 
\begin{equation}
	\delta \Big(u_\x^a \text R^{\y\x}_a\Big) = \text R^{\y\x,\,\e}_A \Big(w_\x^A - \dot\xi^A_\x \Big) = 0\;,
\end{equation}
so that not only do we have $w_\x^a = \dot\xi^a_\x$ but also $w_\x^A = \dot \xi^A_\x$. This then means that we must have\footnote{Here the semicolon is, as usual, a short-hand notation for covariant derivative $A_{;a}= \nabla_aA$.} $u_\e^a\xi_\x^b X^D_{\e\,;ba} = 0$, which in turn implies  that the orthogonality conditions for the viscous stress tensors
\begin{equation}
	S^\x_{ab}u_\x^a = \mathcal{S}^{\x\y}_{ab} u_\x^a = s^{\x\y}_{ab}u_\y^a = 0 \;,
\end{equation}
are automatically satisfied at linear order. 

From \cref{eq:ExpansionViscousTensorsMatter} we can find the expansion for the spacetime viscous tensors through 
\begin{subequations} \label{eq:PertExpViscStressesSpacetime}
\begin{align}
	&\delta S^\x_{ab} = \delta S^\x_{DE} \Psi^D_{\e\,a} \Psi^E_{\e\,b} - S^\x_{DE} \Big( \xi^D_{\x\,,a}\Psi^E_{\e\,b} +\Psi^D_{\e\,a} \xi^E_{\x\,,b}\Big) \;,\\ 
	& \delta s^{\x\y}_{ab} = \delta s^{\x\y}_{DE} \Psi^D_{\e\,a} \Psi^E_{\e\,b} - s^{\x\y}_{DE}\Big( \xi^D_{\y\,,a}\Psi^E_{\e\,b} + \Psi^D_{\e\,a}\xi^E_{\y\,,b}\Big) \;,\\
	&\delta \mathcal{S}^{\x\y}_{ab} = \delta \mathcal{S}^{\x\y}_{DE} \Psi^D_{\e\,a} \Psi^E_{\e\,b} - \mathcal{S}^{\x\y}_{DE}\Big( \xi^D_{\x,a}\Psi^E_{\e\,b} +\Psi^D_{\e\,a} \xi^E_{\y\,,b}\Big) \;,
\end{align}
\end{subequations}
while for the resistive terms associated with $s^{\x\y}_{ab}$ and $\mathcal{S}^{\x\y}_{ab}$ we have 
\begin{subequations}\label{eq:PertReactiveViscousTensor}
\begin{align}
	\delta r^{\x\y}_a &= \frac{1}{2} \delta s^{\x\y}_{DE} \nabla_ag_\e^{DE} - \frac{1}{2} s^{\x\y}_{DE}  \partial_a \big[ g^{bc} (\xi^D_{\y\,,b} \Psi^E_{\e\,c} + \Psi^D_{\e\,b} \xi^E_{\y\,,c})\big] \;,\\
     \delta \mathcal{R}^{\x\y}_a &= \frac{1}{4}\delta \mathcal{S}^{\x\y}_{DE} \nabla_ag^{DE}_\e - \frac{1}{2}\mathcal{S}^{\x\y}_{DE}  g^{bc}\Big(\xi^D_{\x\,,b} \nabla_a \Psi^E_{\e\,c} + \Psi^D_{\e\,b} \nabla_a\xi^E_{\y\,,c}\Big) \;,
\end{align}
\end{subequations}
where we made use of the fact that $[\delta,\nabla_a]=0$ because of $\delta g^{ab}=0$ (see the discussion at the end of \cref{sec:PerturbativeExpansion}). 

Having ``understood'' how we may perturb the terms $R^\x_a$ and $D^\x_{ab}$, let us focus on the remaining pieces of the equation of motion. A quick look back at \cref{eq:DissipativeVariationalEoM} reveals that the only terms we still have to discuss are $\delta\Gamma_\x $ and $\delta \mu^\x_a$. For the particle creation rate we have (see \cref{eq:deltanexa})
\begin{equation}
	\delta \Gamma_\x = \nabla_a \delta n_\x^a = \dot{\delta n_\x} + \nabla_a(\bar n_\x w_\x^a) 
\end{equation}
while for the x-species momentum, we get 
\begin{equation}
	\delta \mu^\x_a = \delta (\mathcal{B}_\x  n_\x) u^\e_b + \bar{\mathcal{B}}_\x \bar n_\x w^\x_b + \sum_{\y\neq \x} \delta(\mathcal{A}_{\x\y} n_\y) u^\e_b + \bar{\mathcal{A}}_{\x\y}\bar n_\y w^\y_b \;.
\end{equation}
Using the fact that we identified $\M_\x$ with $\mu_\x$ we have 
\begin{equation}\label{eq:deltaBarMMx}
\begin{split}
	\delta\M_\x &=   \delta \big(-\mu_\x^a u^\x_a\big) =   -\big(\bar\mu_\x^a  w_\x^a + \delta \mu_\x^a u^\e_a  \big)\\
	& = \delta \Big(\mathcal{B}_\x n_\x + \sum_{\y\neq \x} \mathcal{A}_{\x\y}n_\y\Big) \;,
\end{split}    
\end{equation}
and since  $\B_{\x}$ and $\A_{\x\y}$ are ultimately functions of $n_\x^2$ and $n_{\x\y}^2$, we may use 
\begin{subequations}
\begin{align}
& \delta \mathcal{B}_\x = \Bigg( 2n_\x\frac{\partial \mathcal{B}_\x}{\partial n_\x^2}  + n_\y\frac{\partial \mathcal{B}_\x}{\partial n_{\x\y}^2} \Bigg) \delta n_\x +  \Bigg( 2n_\y\frac{\partial \mathcal{B}_\x}{\partial n_\y^2}  + n_\x\frac{\partial \mathcal{B}_\x}{\partial n_{\x\y}^2} \Bigg) \delta n_\y \;,\\
& \delta \mathcal{A}_{\x\y} = \Bigg( 2n_\x\frac{\partial \mathcal{A}_{\x\y}}{\partial n_\x^2}  + n_\y\frac{\partial \mathcal{A}_{\x\y}}{\partial n_{\x\y}^2} \Bigg) \delta n_\x +  \Bigg( 2n_\y\frac{\partial \mathcal{A}_{\x\y}}{\partial n_\y^2}  + n_\x\frac{\partial \mathcal{A}_{\x\y}}{\partial n_{\x\y}^2} \Bigg) \delta n_\y \;,
\end{align}
\end{subequations}
in \cref{eq:deltaBarMMx}. This way, making use of definitions in \cref{eq:SecondOrderMasterUsefulDef}, we arrive at
\begin{equation}\label{eq:PertBarMMxFico}
	\delta \M_\x = \big(\bar\B_\x c_\x^2 - \bar\A^{\x\x}_{uu}\big)\delta n_\x -\big(\bar\A^{\n\s}_{uu} +\bar\chi ^{\n\s}_{uu}\big)\delta n_\y \;,
\end{equation}
and we see that the parameters that enter the dissipative fluid equations are the entrainment coefficients (and  first derivatives; that is, second order derivatives of $\Lambda(n_\x^2, n_{\x\y}^2)$) and the (up to second order) derivatives of the function $\N_\x(X_\x,\,X_\y,\,g^{AB}_\x,\,g^{AB}_\y,\,g^{AB}_{\x\y})$.  

Having outlined the perturbative framework, it is natural to ask how many dissipative channels the (general) model contains. Or, to be more specific, how many ``dissipation coefficients'' would have to be determined from microphysics? According to the expansion scheme we have developed so far, the perturbative expressions for the dissipative terms will ultimately involve all second and first order derivatives of the $\N_\x^d$ when considered as functions of $X_\x,\,X_\y,\,g^{AB}_\x,\,g^{AB}_\y,\,g^{AB}_{\x\y}$. Also, to make use of the model we need to specify the entrainment coefficients and their derivatives in the combinations from \cref{eq:SecondOrderMasterUsefulDef}. This means that the most general model one can think of contains a large number of coefficients. However, they should be, in general, known once a specific model is chosen; that is, once the explicit functional forms of $\Lambda$ and the $\N_\x^d$ have been provided. For example, if nuclear physics calculations are used to determine these explicit forms, they must be done in such a way that the constraints which arise from requiring a meaningful equilibrium configuration are taken into account, and they must ensure that the second law of thermodynamics is obeyed. If Onsager-type reasoning \cite{onsager31:_symmetry} is invoked to ensure $\Gamma_\s$ is positive (up to second order), then explicit use of
\begin{equation}\label{eq:GeneralEntropyProductionRate}
	T \Gamma_\s = - D_{ba}^\s \nabla^bu_\s^a - u_\s^a \text R^\s_a  \; ,
\end{equation}
where $T = - u^a_\s \mu^\s_a$ is the temperature, would have to be made.

\section{Model comparison}\label{sec:StdQuantities} 

As an intuitive application of the formulation it is useful to make contact with existing models for general relativistic dissipative fluids, in particular, the classic work of Landau-Lifschitz and Eckart and the second-order M\"uller-Israel-Stewart model. Specifically, we want to understand how  standard quantities (like shear and bulk viscosity) enter the present formalism. Therefore, we need to see if the dissipative terms of the existing models can be matched with terms in the action-based description. This procedure is fairly straightforward. 

The action-based model provides the total fluid stress-energy-momentum tensor, so  we only have to decompose it in the usual way:
\begin{equation}
	T^{ab} = (\bar p+\chi)\perp^{ab} + \varepsilon u^a u^b + 2q^{(a}u^{b)} + \chi^{ab} \;,
\end{equation}
where $q^a$ is orthogonal to $u^a$ while  $\chi^{ab}$ is traceless and transverse (w.r.t to $u^a$) as usual.
In this expression, the fluxes are defined with respect to some observer with four-velocity $u^a$. In order to be consistent with the perturbative expansion outlined above, we take this observer to be associated with the thermodynamical equilibrium, i.e.~$u^a = u^a_\e$.

\subsection{Equating the Flux Currents}\label{sec:FluxCurrents}

Let us first consider  the heat. We can read off the heat flux from the total stress-energy-momentum tensor as 
\begin{equation}
	q^a = -\varepsilon u^a_\e - T^{ab} u^\e_b = - \perp ^a_b T^{bc}u^\e_c \;.
\end{equation}
First, let us note that there is no contribution\footnote{At linear order.} coming from the dissipative part of the stress-energy-momentum tensor $D^{ab}$. In fact, making use of \cref{eq:mixedviscousdie,eq:viscousdie,eq:PertExpViscStressesSpacetime}, it is easy to show that $D^{ab}u^\e_a = (\delta D^{ab})u^\e_a = \mathcal{O}(2)$. Let us therefore consider the non-dissipative part of $T^{ab}$. For the generalized pressure we have to first order  
\begin{equation}
	\Psi = \Lambda + \sum_{x} n_\x\mu_\x = -\bar\varepsilon_\e + \bar\mu \bar n + \bar T \bar s + \sum_{x=n,s} \bar n_\x \delta \mu_\x = \bar p +  \sum_{x=n,s} \bar n_\x \delta \mu_\x 
\end{equation}
where we have used the minimum energy condition (\cref{eq:MinimumeEnergyCondition}) and the equilibrium Euler relation. Using 
\begin{equation}
	\sum_\x n_\x^a\hat\mu_\x = \sum_\x n_\x^a\mu_\x + \mathcal{O}(2) \approx \sum_\x \Big[\bar n_\x\bar\mu_\x u^a_\e +\bar n_\x\delta\mu_\x u^a_\e +\bar\mu_\x\big(\delta n_\x u^a_\e + \bar n_\x w_\x^a\big)\Big] 
\end{equation}
we then identify the heat flux as
\begin{equation}\label{eq:modelheatflux}
	q^a = \sum_\x \bar\mu_\x \delta n_\x^a = \bar\mu \bar n \,w_\n^a +\bar T\bar s \,w_\s^a \ . 
\end{equation}
Here we have repeatedly used the Euler relation and the minimum energy condition \cref{eq:MinimumeEnergyCondition}. We note that this quantity is consistent with the definition used in the classic models, see \citep{NilsGregReview}. 

Let us now move on to the other fluxes and, as before, first focus on the non-dissipative contribution. It is easy to check that
\begin{equation}
\begin{split}
	T^{ab}_{\n.\d.} = &\bigg(\bar p + \sum_\x \bar n_\x \delta \mu_\x\bigg)g^{ab} + (\bar p+\bar\varepsilon_{\e})u^a_\e u^b_\e \\
    &+\sum_\x \Big[\bar\mu_\x\bar n_\x u^b_\e w_\x^a + \bar n_\x u^a_\e \big(\delta\mu_\x u^b_\e + \B_\x\bar n_\x w_\x^b + \sum_{\y\neq \x} \A_{\x\y} \bar n_\y w_\y^b\big)\Big]\;,
\end{split}
\end{equation}
so that, using the standard decomposition above one arrives at  
\begin{equation} 
(\bar p+\chi)\perp^{ab} + \chi^{ab} = \perp^a_c\perp^b_dT^{cd}=T^{ab} +T^{ad} u^\e_d u^b_\e + T^{cb} u^\e_c u^a_\e + \varepsilon u^a_\e u^b_\e \;.
\end{equation}
If we now use  the non-dissipative contribution $T^{ab}_{\n.\d.}$ in this  equation, we get
\begin{equation}
	(\bar p+\chi)\perp^{ab} + \chi^{ab} = (\bar p + \sum_\x \bar n_\x \, \delta \mu_\x)\perp^{ab} = (\bar p + \delta \Psi) \perp^{ab} \;.
\end{equation}
That is, there may be a first-order correction in the pressure coming from $T^{ab}_{\n.\d.}$. Next, let us consider the contribution due to the non-dissipative part. From \cref{eq:PertExpViscStressesSpacetime} we see that %
\begin{equation}
	\perp^a_c\perp^b_d D^{cd}= D^{ab}=\delta D^{ab} \;.
\end{equation}
Putting everything together, we have identified
\begin{subequations}
\begin{align}
	\hat\chi &= \delta \Psi + \frac{1}{3} g_{ab}\delta D^{ab} \;,\\
	\hat\chi^{ab} &= \delta D^{\langle ab \rangle} \;, \\
	\hat q^a &= \bar\mu\bar n w_\n^a + \bar T\bar s w_\s^a \;,
\end{align}
\end{subequations}
where we reintroduced the ``hat'' to stress that these fluxes are measured by the equilibrium observer while the angle brackets mean that we are taking the trace-free symmetric part of the tensor. 

\subsection{Example: A single viscous fluid}\label{sec:ViscousFluid}
We now consider the specific example of a two-component, single viscous fluid. The two species are matter, with non-equilibrium flux $n^a = n u^a_\f$, and entropy, with non-equilibrium flux $s^a = s u^a_\f$. In this simple case, we assume that the non-equilibrium fluxes remain parallel\footnote{We note that a ``real'' two-fluid model would involve two independent fluid degrees of freedom $n^a_\n,\,n^a_\s$. By forcing them to move together we are imposing quite strong constraints on the model. Basically, we are assuming that the timescale over which the entropy current relaxes to the particle flow is short enough that it may be neglected.}, meaning $w^a_\n = w^a_\s = w^a$ and therefore
\begin{equation}
\begin{split}
    n_\n^a = n u^a_\f = n(u^a_\e + w^a) \;,\\
    n_\s^a = s u^a_\f = s(u^a_\e + w^a) \;,
\end{split}
\end{equation}
where again $u^a_\e$ is the equilibrium flow. In this case we do not have resistive terms because the two fluids are locked together from the beginning. Dissipation enters by assuming both currents depend on the (single) projected metric
\begin{equation}
\begin{split}
	\N_\n = \N_\n(X^A,\, g^{AB}) \;, \\
    \N_\s = \N_\s(X^A,\, g^{AB}) \;. \\
\end{split}
\end{equation}
In practice, this means that we will have additional terms due to $S^\s_{ab}$ and $S^\n_{ab}$ in the equations of motion.

Also, the creation rate $\Gamma_\n$ has to vanish\footnote{The matter particle flux $n_\n^a$ is conserved as it is identified with the baryon current.}; this implies
\begin{equation}
	\Gamma_\n = -\frac{1}{\mu_\n} S^\n_{ab} \nabla^a u^b_\f = 0 \Longrightarrow S^\n_{ab} = 0 \;,
\end{equation}
as, by construction, the viscous stress tensor $S^\n_{ab}$ must be orthogonal to $u_\f^a$ (see \cref{eq:StdViscousTensorSlim} and \cite{2015CQAnderssonComer} for further details).
As a result, the final form of the non-linear equation of motion is 
\begin{equation}\label{eq:ViscousOneFLuidEq}
	2n_\n^a\nabla_{[a}\mu^\n_{b]} + 2n_\s^a\nabla_{[a}\mu^\s_{b]} + \Gamma_\s \mu_b^\s = - \nabla^a S^\s_{ab} \;.
\end{equation}
Note that, when we linearize, the term involving $\Gamma_\s$ will not appear in the equations, because $\Gamma_\s$ has no linear contributions---entropy is expanded around a maximum, leaving only second-order terms.

Our next step is to use the expansion formalism developed in the previous sections to determine the explicit form of the viscous stress tensor $S^\s_{ab}$.
Let us start by considering the equilibrium (minimum energy) conditions. Clearly, we should have
\begin{equation}
	S^{\s,\,\e}_{AB} = 2\M_\s \frac{\partial \N_\s^\d}{\partial g^{AB}} = 0 \Longrightarrow \frac{\partial \N_\s^\d}{\partial g^{AB}} = 0 \;.
\end{equation}
It also makes sense to assume $\partial \N_\s^d/\partial X^A = 0$. To see why, let us forget for the moment that the two species are locked together and consider:
\begin{equation}
\begin{split}
	\bar R^\s_A &= \M_\n \frac{\partial \N_\n^\d}{\partial X^A_\s} - \M_\s \frac{\partial \N_\s^\d}{\partial X^A_\n} = -\M_\s \frac{\partial \N_\s^\d}{\partial X^A_\s} - \M_\s \frac{\partial \N_\s^\d}{\partial X^A_\n}  \\
    &= - 2\M_\s \frac{\partial \N_\s^\d}{\partial X^A_\s} =  - 2\M_\s \frac{\partial \N_\s^\d}{\partial X^A} = 0 \;,
\end{split}
\end{equation}
where we initially distinguished between the two constituents' matter-space coordinates, and used the equilibrium condition. The condition $\partial \N_\x^\d/\partial X^A = 0$ is thus motivated by the fact that the resistive term vanishes (because the two currents are effectively locked).

As a result of these constraints we have $\delta \N_\x^\d = \mathcal{O}(2)$, $\delta \Psi = \mathcal{O}(2)$ and the viscous stress tensor becomes (see \cref{eq:ExpansionViscousTensorsMatter,eq:PertExpViscStressesSpacetime}) 
\begin{equation}\label{eq:ViscousFluidSAB}
\begin{split}
	\delta S^\s_{ab} &= \Bigg[ 2\bar\M_\s \delta \bigg(\frac{\partial \N_\s^\d}{\partial g^{AB}}\bigg) \Bigg]\Psi^A_{\e\,a}\Psi^B_{\e\,b} \\
&= 2\bar T \Bigg[ \frac{\partial \N_\s^\d}{\partial X^C \partial g^{AB}}\delta X^C + 2 \frac{\partial \N_\s^\d}{\partial g^{DE} \partial g^{AB}}\delta g^{DE}  \Bigg]\Psi^A_{\e\,a}\Psi^B_{\e\,b} \;,
\end{split}
\end{equation}
Let us now rewrite the terms within the square brackets as 
\begin{equation}
\begin{split}
	A_{CAB} &=  2\frac{\partial \N_\s^d}{\partial X^C \partial g^{AB}} \;,\\
    \Sigma_{DEAB} &= 4 \frac{\partial \N_\s^d}{\partial g^{DE} \partial g^{AB}} \;.
\end{split}
\end{equation}
We further assume that $A_{CAB}$ is zero as this involves degrees of freedom we do not need to recover Navier-Stokes equations.

Before moving on with the model specification, let us introduce a slight generalization to the original model from \cite{2015CQAnderssonComer}. The idea is to take the normalizations $\N_\x^d$ as functionals---instead of functions---of the additional variables. This does not constitute a major difference as the equations of motion, and particle production rate formulae, remain unchanged. Still, the step can be taken subject to the following caveat: The  (functional) integration should extend (at most) to the spacetime region that is causally connected with each point. Here, we will assume that the analysis is done locally in space but not necessarily in time, i.e. on the world-tube formed by the spatial part of the region $\delta {\cal M}$ in \cref{fig:FluidElem}.  
In the present example this would mean 
\begin{equation}
\begin{split}
	\N_\x^d[g^{AB}] = &\N_\x^d[g^{AB}_\e] + \int  \frac{\delta \N_\x^d}{\delta g^{AB}(x)}\delta g^{AB}(x) \text{d}^4x+ \\
    &+\frac{1}{2}\int \frac{\delta ^2 \N_\x^d}{\delta g^{AB}(x)\delta g^{CD}(y)}  \delta g^{AB}(x)\delta g^{CD}(y)\text{d}^4x\text{d}^4y 
\end{split}
\end{equation}
where the first two terms vanish because (i) $\N^d_\x$ vanishes at equilibrium, and (ii) the minimum energy condition. 
The key step then is to replace the ordinary partial derivatives with functional derivatives in the various expressions we have discussed, so that the viscous stress tensor will be
\begin{equation}
	S^\s_{AB}(x) =  2 \bar T \delta \Bigg( \frac{\delta \N_\s^d}{\delta g^{AB}} \Bigg)(x) = 2 \bar T \int  \frac{\delta^2 \N_\s^d}{\delta g^{AB}(x)\delta g^{CD}(y)}\delta g^{CD}(y) \text{d}^4y \;.
\end{equation}
We can now formally introduce a set of spatial coordinates $\bar x$ comoving with the equilibrium observer and attached to the world-tube, and take the time coordinate to be the equilibrium worldline's proper time $\tau$. Also, to enforce locality in space we let
\begin{equation}
	 \frac{\delta^2 \N_\s^d}{\delta g^{AB}(x)\delta g^{CD}(y)} = \frac{1}{4}\Sigma_{ABCD}(\bar x, \tau_x-\tau_y) \, \delta^3(\bar x - \bar y)  \; .
\end{equation}
where the causality condition $\tau_x - \tau_y \ge 0$ is assumed to be encoded within $\Sigma_{ABCD}$.  

Let us first of all, as a consistency check, show that the formula for the particle production rate remains unaltered by these modifications. We have (see \cref{eq:NonConservativeGamma})
\begin{equation}
\begin{split}
	\mu_\x\Gamma_\x &= \frac{1}{3!} \mu_\x^{ABC}\frac{d n^\x_{ABC}}{d\tau_\x} = \bar\M_\x \frac{d}{d\tau_\x}\big(\bar\N_\x^c + \bar\N_\x^d\big) \\
    &= \M_\x \Bigg( \frac{d\N_\x^d}{d\tau_\x} + \frac{1}{2}\N_\x^dg^\x_{AB}\frac{dg_\x^{AB}}{d\tau_\x}\Bigg) \;.
\end{split}
\end{equation}
where, for a single viscous fluid the result simplifies to 
\begin{equation}
	\Gamma_\x =  \frac{d\N_\x^d}{d\tau} +  \mathcal{O}(3) \;.
\end{equation}
If in particular we consider the entropy production rate $\Gamma_\s$ we have 
\begin{equation}
\begin{split}
	\N_\s^d &= \frac{1}{8} \int  \Sigma_{ABCD}(\bar x,\tau-\tau') \delta g^{AB}(\bar x,\tau) \delta g^{CD}(\bar x,\tau') \text{d}^3\bar x\, \text{d}\tau \,\text{d}\tau'
\end{split}
\end{equation}
To compute the entropy creation rate we have to use the chain rule (generalized to functionals) on $\N_\s^d[g^{AB}(x)]$
\begin{equation}
	\frac{d\N_\s^d}{d\tau} = \int  \frac{\delta N_\s^d}{\delta g^{AB}(y)} \frac{\delta g^{AB}(y)}{\delta \tau(x)} \text{d}^4y\;.
\end{equation}
But, because $g^{AB}$ is a ``normal'' function of the spacetime coordinates 
\begin{equation}
	\frac{\delta g^{AB}(y)}{\delta \tau(x)} = -2\delta^4(x-y) D^{(A}w^{B)} (x)
\end{equation}
so that we are left with 
\begin{equation}
\begin{split}
	\Gamma_\s &= -\frac{1}{\bar T} S_{AB}(\bar x,\tau) D^{(A}w^{B)} (\bar x,\tau) \;. 
\end{split}
\end{equation}

We can now make use of this ``functional generalization'' to recover the Navier-Stokes model for a bulk- and shear viscous fluid. To focus on the important parts, let us consider first the purely bulk-viscous case
\begin{equation}\label{eq:PurelyBulkExample}
	S(\tau) = \int K(\tau-\tau') A(\tau') \text{d}\tau'
\end{equation}
where $S$ represents the trace of the viscous-stress tensor while A stands for the trace $\text{tr }\delta g^{AB}$. Because of the difference between $\delta g^{AB}$ and $\dot g^{AB}$---the former involves gradients in the displacement while the latter depends on the velocity---in order to recover a Navier-Stokes model we need to take 
\begin{equation}
	K(\tau-\tau') = -T\zeta\partial_{\tau'}\delta (\tau-\tau') \;,
\end{equation}
since this would give
\begin{equation}
	S(\tau) = T\zeta\int [-\partial_{\tau'}\delta (\tau-\tau')] A(\tau')\text{d}\tau' =T\zeta \int\delta (\tau-\tau') \partial_{\tau'}A(\tau')\text{d}\tau' = T\zeta\frac{dA}{\text{d}\tau} \;,
\end{equation}
as desired. We now implement this for the bulk- and shear model. We can do this using the standard decomposition of the bulk and shear response as
\begin{equation}\label{eq:SigmaDecomposition}
	\Sigma_{ABCD} = \Sigma_{ABCD}^\b + \Sigma_{ABCD}^\s \;,
\end{equation}
with 
\begin{equation}\label{eq:functionalSigmas}
\begin{split}
	\Sigma_{ABCD}^\b &=  \frac{\zeta(x)}{\bar T}\, g^\e_{AB}g^\e_{CD} \delta^3(\bar x - \bar y) \, q_\b(\tau_x-\tau_y)\;,\\
    \Sigma_{ABCD}^\s &= 2\frac{\eta(x)}{\bar T}\, \bigg( g^\e_{A(C}g^\e_{B)D} - \frac{2}{3}g^\e_{AB}g^\e_{CD}  \bigg )\delta^3(\bar x - \bar y) \, q_\s(\tau_x-\tau_y) \;,
\end{split}
\end{equation}
where the two kernels would be\footnote{Note that we have chosen to separate the bulk- and shear channels as usual, even though the present construction allows for anisotropic response in the velocity gradients to viscosity relation. We have also introduced two independent kernels to allow for different response to bulk and shear strain rates.} the same  $q_\b = q_\s = - \partial_{\tau_y}\delta(\tau_x-\tau_y)$. It then follows that the only viscosity tensor of the model is
\begin{equation}
	 S^\s_{ab} = \chi_{ab} + \chi \perp_{ab} = \frac{1}{3}\zeta \, \theta \perp_{ab} + \eta\,\sigma_{ab} \;.
\end{equation}
It is also easy to check that we can enforce compatibility with the second law by fixing the sign of the bulk-and shear viscosity coefficients. In fact,
\begin{equation}
	\Gamma_s = \frac{\zeta}{\bar T}\, \theta^2 + \frac{\eta}{\bar T}\, \sigma^{ab} \sigma_{ab} \ge 0 \;.
\end{equation}%
where $\zeta,\eta \ge 0$. With these relations we have recovered the usual relativistic Navier-Stokes equations (the Landau-Lifschitz-Eckart model for a viscous fluid).

Let us conclude by pointing out that, to write down the full set of equations one should also expand the ``Euler part'' of the equation of motion, i.e. the left-hand-side of \cref{eq:ViscousOneFLuidEq}. We have provided all the ingredients necessary for the explicit calculation, but leave it out here as the result is not new and not particularly relevant for the present discussion (see \cite{NilsGregReview} for further details).

\section{Cattaneo-type equations}
\label{sec:Cattaneo}

As a practical example of the first-order expansion we outlined the model for a single (bulk and shear) viscous fluid, and showed how this leads to the expected form of the relativistic Navier-Stokes equations. The derivation shows that the action-based formalism encodes the previous models. It is also clear that the formalism allows us to consider much more complicated settings, should we need to do so. However, the discussion of the first-order results is clearly not complete, because the final set of equations is widely known to suffer from causality/stability issues. In fact, the work of \citet{HiscockLindblom1storder} has shown these first-order models to be generically unstable. In practice, this means that if we set the system to deviate slightly from an equilibrium state, the deviations may grow exponentially and eventually diverge. Conversely, the second-order M\"ueller-Israel-Stewart theories have been shown to satisfy the conditions for stability and causality (see \cite{HiscockLindblom1983}) in the linear regime we are considering here. This has led to the common belief that all possible first-order theories are unstable and acausal, and that for these issues to be resolved one has to work at second order. 

The issue has recently drawn attention because of the relevance of (general relativistic) dissipative fluid models in the new gravitational-wave era and for the modelling of heavy-ion collisions. In particular, in recently proposed field-theory-based models (see \cite{Kovtun19, Bemfica19}) one postulates that the thermodynamical fluxes can be expanded in terms of the usual hydrodynamic variables and their derivatives. Halting the derivative expansion at first order and performing a stability analysis, the authors of \cite{Kovtun19,Bemfica19} showed that there exists a consistent set of constraints on the expansion parameters such that these models pass some of the stability and causality conditions. 

A recent analysis by \citet{Lory1stInsta} shed new light on this matter,  showing that Landau-Lifschitz-Eckart model instabilities are due to the enforcement of the second law on an entropy function that is not maximized at equilibrium, while the field-theory-based models can be made stable by allowing for small violations of the second law. 

Another important aspect of the problem is provided by \cite{Lopez2011,NilsHeat2011}, where it is demonstrated that, for a fluid model with heat-flux,  one can resolve the stability/causality issues at first order by properly accounting for the entrainment between matter and entropy currents---retaining the compatibility with the second law. 
We will now discuss these issues using the single viscous fluid as a case study. 

Essentially, the problem must be addressed in a different way, as the key ingredient used to solve the heat-flux case (see \cite{Lopez2011,NilsHeat2011}) accounting for the entropy inertia, will not work for the present case as our model setting does not involve relative flows. Instead, we will sketch how one can obtain---taking the action-based formalism as the  starting point---a set of equations that is consistent with linearizing the equations of M\"uller-Israel-Stewart. The argument is similar to that of Rational Thermodynamics, which stresses the importance of the ``principle of memory or heredity'' (see \cite{EIT}). 

Let us first focus on the bulk viscosity case, and then extend the results to the bulk- and shear viscous model. Recalling \cref{eq:PurelyBulkExample}, the first step would again be to assume
\begin{equation}
	K(\tau-\tau') = -\partial_{\tau'}g (\tau-\tau')
\end{equation}
so that 
\begin{equation}
	S(\tau)= \int K(\tau-\tau') A(\tau')\text{d}\tau' = \int g(\tau-\tau') \frac{dA(\tau') }{\text{d}\tau'}\text{d}\tau'  \;.
\end{equation}
We can then look for the convolution kernel $g$ such that the bulk-viscous scalar $S$ satisfies an equation of the Cattaneo type
\begin{equation}
	t_\b \dot S = -S -\zeta \frac{dA}{d\tau} \;,
\end{equation}
where $t_\b$ is the relaxation time-scale of the bulk-viscosity response. In terms of the convolution kernel $g$ this would mean
\begin{equation}
	t_\b\partial_\tau g(\tau-\tau') + g(\tau-\tau') = -\zeta\delta(\tau-\tau')\;.
\end{equation}
Solving this equation we get 
\begin{equation}
	g(\tau-\tau') = -\frac{\zeta}{t_\b} e^{-(\tau-\tau')/t_\b}\theta(\tau-\tau') \;,
\end{equation}
and one can check this is the correct result by direct computation:
\begin{equation}
\begin{split}
	\partial_\tau  g(\tau-\tau')&= \frac{1}{t_\b}\frac{\zeta}{t_\b}e^{-(\tau-\tau')/t_\b}\theta(\tau-\tau') -\frac{\zeta}{t_\b}  e^{-(\tau-\tau')/t_\b} \delta(\tau-\tau') \\
	&= -\frac{1}{t_\b} g(\tau-\tau') - \frac{\zeta}{t_\b} \delta(\tau-\tau') \;.
\end{split}
\end{equation} 

Intuitively, one would like to recover the  Navier-Stokes response in the limit of very short relaxation timescales $t_\b\to 0$. It is clear that to do so, we need to let the bulk viscosity coefficient diverge in the short timescale limit. That is, we need  
\begin{equation}
	\zeta(x)e^{-(\tau - \tau')/t_\b} \underset{t_\b\to 0}{\longrightarrow} \zeta_{NS} \;,
\end{equation}
where $\zeta_{NS}$ is the usual Navier-Stokes bulk viscosity coefficient. This is in accord with the parabolic limit of \cite{BulkLory}.

We are now ready to go back to the full bulk- and shear viscous model.  We will retain the structure and symmetries from before (see \cref{eq:functionalSigmas,eq:SigmaDecomposition}), but introduce two different convolutions $q_\b$ and $q_\s$ to account for retarded response to bulk and shear strain rates. In essence, we have shown how we can implement a retarded response of the Cattaneo type in the action-based model by assuming that $S_{AB}$ (and therefore $D_{ab}$ as well) is an integral function of $g^{AB}$. The question then is, does this mean that the final fluid equations are integro-differential equations? Fortunately the answer is no. In fact, we have shown that, by a suitable choice of the response function $q(\tau-\tau')$, the fluxes satisfy an equation of the Cattaneo-type. Therefore, instead of solving an integro-differential equation, one should treat $S^\s_{ab}= \chi \perp_{ab} + \chi_{ab} $ as an unknown in \cref{eq:ViscousOneFLuidEq}, and add two equations to the system
\begin{subequations}\label{eq:telegraph-typeEquations}
\begin{align}
	\chi + t_\b\,\dot \chi &= - \zeta \,\theta\;, \\
    \chi_{ab} + t_\s\,\dot\chi_{ab} &= -\eta\,\sigma_{ab} \;.		
\end{align}
\end{subequations}

This means that, at the end of the day, to actually solve a set of differential dissipative equations at first order, we have to treat the fluxes as additional unknowns, for which one has to provide equations that are not given by the stress-energy-momentum conservation law $\nabla_aT^{ab}=0$. This is reminiscent of the EIT paradigm, where one postulates from the beginning an entropy function that depends on an additional set of quantities---the thermodynamical fluxes. The difference is that the microphysical origin of the equation for the fluxes is now more clear. It is also worth noting that equations of the Cattaneo-type for the fluxes cannot be obtained in the field-theory-based models, as the constitutive equations are given in terms of the usual equilibrium variables (like $\mu,\,T$) and their derivatives---so that terms with derivatives of the fluxes (like $\dot\chi$) do not appear. 

\Cref{eq:telegraph-typeEquations} is (formally) the same as in the linearized version of M\"ueller-Israel-Stewart model, which has been shown to be stable and causal. In theory, nothing prevents us from choosing a different form for the retarded response $q$ which could  lead to acausal/unstable behaviour. However, the form of $q$ suggested above has a clear physical interpretation and is microphysically motivated. If one wants to come up with an alternative, this would need to be motivated by microphysical arguments, as well. 

Let us now consider the implications of the Cattaneo laws for the entropy production rate. As in the Navier-Stokes model sketched above we have 
\begin{equation}\label{eq:BulkShearEntropy}
\begin{split}
	\Gamma_\s &= \int  \Sigma_{ABCD}(x,\tau-\tau') D^{(C}w^{D)}(\bar x,\tau')D^{(A}w^{B)}(\bar x,\tau) \text{d}\tau'= \\
    &= -\frac{1}{\bar T} S_{AB}(\bar x,\tau) D^{(A}w^{B)} (\bar x,\tau)
\end{split}
\end{equation}
Again, let us use the bulk-viscous case to highlight the relevant features. We then have 
\begin{equation}\label{eq:CITentropy}
	\Gamma_\s = -\frac{\zeta}{t_\b} \int_{-\infty}^\tau  e^{-(\tau-\tau')/t_\b}\theta(\tau)\theta(\tau') \,\text{d}\tau'
\end{equation}
where $\theta$ is the expansion rate. It is clear that, because the expansion rate is evaluated at different times, we cannot guarantee the positivity of the entropy production rate just by fixing the sign of the bulk-viscosity coefficient as in the previous Navier-Stokes case. However, we will argue this result is not as dramatic as it may appear at first sight---at least not in the regimes relevant for physical predictions. Because of the exponential in the integral, we can assume that the values for the expansion at times $\tau'$ can be neglected for $\tau'$ that is a few $t_\b$ away from $\tau$. As a result we can expand $\theta(\tau')$ as 
\begin{equation}
	\theta(\tau') = \theta(\tau) + \frac{d\theta(\tau)}{\text{d}\tau} (\tau'-\tau) = \theta(\tau) + \dot\theta(\tau)(\tau'-\tau)
\end{equation}
so that we obtain
\begin{equation}\label{eq:CITentropySimplified}
	\Gamma_\s = -\zeta \big[\theta^2(\tau) -t_\b \theta(\tau)\dot\theta(\tau)\big] \;.
\end{equation}
Again, because of the product of the expansion rate with its time derivative, the entropy production rate cannot be made generally positive simply by fixing the sign of the bulk-viscosity coefficient $\zeta$. However, as shown in \cite{Geroch1995,Lindblom1995}, physical fluid states relax---on a timescale characteristic of the microscopic particle interactions---to ones that are essentially indistinguishable from the simple relativistic Navier-Stokes description. Translated to the present context this would mean 
\begin{equation}
	t_\b \dot\theta(\tau) \approx t_\b \frac{\Delta \theta}{\tau_{\text{hydro}}} \to 0
\end{equation}
because the ratio between the relaxation timescale $t_\b$ and the hydrodynamical timescale $\tau_{\text{hydro}}$ is effectively negligible. As a result, for actual physical applications we can neglect the second term in the entropy production rate and compatibility with thermodynamic second law is restored. In a way, this would mean that for bulk- and shear viscous fluids, we can introduce Cattaneo laws for the fluxes---to fix the causality and stability issues of the Navier-Stokes model---while the physical content will be precisely that of Navier-Stokes. 
\section{Conclusions and final remarks}

We have considered the close-to-equilibrium regime of the action-based model of \citet{2015CQAnderssonComer} for dissipative multi-fluid systems. In particular, we have shown that, starting from a set of fully non-linear dynamical equations with only the fluxes as the degrees of freedom, an expansion with respect to (a self-consistently defined) equilibrium can be introduced in a clear fashion, with the line of reasoning being similar to that of usual hydrodynamical perturbation theory. 

After discussing the aspects of equilibrium which can be inferred from the action-based model itself, we established how to construct an expansion in deviations away from equilibrium in a general setting, so that the framework is of wider relevance. In the process we demonstrated the importance of the frame-of-reference of the equilibrium observer. We also noted that the construction promotes the role of the matter space: Instead of it being a mathematical ``trick'' to facilitate a constrained variation, it might well be the arena where the microphysical details are encoded. This is a novel perspective that needs further discussion and consideration.

We then focused on a particular first-order viscous fluid model, with shear- and bulk-viscosity, paying particular attention to the key causality issues. We showed that causal behaviour can be linked to a retarded response function that keeps track of a system's history. The specific form of the response function can be modelled in a phenomenological way---as we did---but should ideally be provided by specific microphysical calculations, for instance by means of the fluctuation-dissipation theorem (see \cite{Reichl} for a general discussion and \cite{EIT} for comments on its role from the EIT perspective).\footnote{We also note that there have been recent efforts to make explicit use of the fluctuation-dissipation theorem to compute response coefficients through Green-Kubo-like formulae, see \cite{MontenegroTorrierCausal,MontenegroTorrierKubo}---although in a special relativistic setting. Nevertheless, there are similarities with the way we deal with causality issues here.} In a sense, the action-based model provides the ``context'', determining the geometric structure and form of the equations of motion, while the detailed microphysics is encoded in the specific response function. 

Building the first-order expansion we made this connection clear, and showed how and where the microphysics enters the discussion. 
An interesting outcome of this analysis is that we showed that there is no need to go to second order in deviations from equilibrium to implement a causal response in the model. This has already been demonstrated for the heat-flux problem (see \citep{NilsHeat2011,Lopez2011}), where the  Cattaneo-type equation for the heat flux is ultimately related to the multi-fluid nature of the problem. The entrainment effect (through which the entropy current gains an effective mass \citep{andersson10:_caus_heat}) results in an inertial heat response.  The case of a single viscous fluid is  different since its retarded response cannot be associated with the multifluid nature of the problem. 

As the variational model is designed for dealing with multi-fluids, the route to further extensions is---at least at the formal level---quite clear.
A natural next step would be the modelling of a viscous fluid allowing for the heat to flow differently from the matter. This application should be fairly straightforward since the two main issues of the problem have now been studied separately. A more challenging  step will be the inclusion of superfluidity. The presence of currents that persist for very long times changes drastically the non-dissipative limit. The model would require the use of more than one equilibrium worldline congruence \cite{2015CQAnderssonComer,ThermoGavassino}, one for each ``superfluid condensate'' and one for all the remaining constituents. We plan to investigate these issues---and the connection to neutron star astrophysics---at a later date.

\acknowledgements

NA acknowledges support from STFC via grant no.~ST/R00045X/1.

\bibliography{action1st.bib}

\end{document}